\documentclass[11pt]{article}
\pdfoutput=1
\usepackage{jheppub}
\usepackage{feynman}
\usepackage[T1]{fontenc} 
\usepackage{pgf,tikz}
\usepackage{mathrsfs}
\usetikzlibrary{arrows}
\usepackage{mathtools}

\title{Integrability of Black Hole Orbits in Maximal Supergravity}

\author[a]{Simon Caron-Huot}
\author[a]{Zahra Zahraee}
\affiliation[a]{Department of Physics, McGill University, 3600 Rue University, Montr\'eal, QC Canada H3A 2T8}
\emailAdd{schuot@physics.mcgill.ca}
\emailAdd{zr.zahraee@physics.mcgill.ca}

\abstract{
We study the dynamics of a pair of extremal (half-BPS) black holes in $\mathcal{N}=8$ supergravity,
as a potentially solvable model of gravitational dynamics.
As a diagnosis of hidden symmetries, we ask whether the perihelion of the orbits precesses over time.
We consider black hole charge vectors with arbitrary misalignment.
First, we use scattering amplitude methods to compute the leading post-Newtonian correction for general mass ratios.
This computation is greatly simplified by introducing a suitable on-shell superspace. 
Second, we study the probe limit to all orders in velocity and Newton's constant
through a ten-dimensional brane setup.
In all cases we find no precession. We relate this to the absence of scalar triangle integrals.}

\begin{document} 
\maketitle
\flushbottom

\def\N{\mathcal{N}{=}8}
\def\Mplanck{M_{\rm pl}}

\def\be{\begin{equation}}
\def\ee{\end{equation}}
\def\ba{\begin{eqnarray}}
\def\ea{\end{eqnarray}}
\def\nl{\nonumber\\}
\def\ket#1{\big| #1\big\rangle}
\def\l{\langle}
\def\r{\rangle}
\def\MM{\mathcal{M}}
\def\AA{\mathcal{A}}
\def\draftnote#1{{\bf [#1]}}
\def\phib{\bar{\phi}}
\def\eps{\epsilon}
\def\bs{\setminus}
\section{Introduction}

There has been a longstanding
interest in applying the methods of quantum field theory and effective field theory to gravitationally bound systems --
for reviews and some recent developments see
\cite{Donoghue:1994dn,Burgess:2003jk,Porto:2016pyg,Buonanno:1998gg,Goldberger:2004jt,Foffa:2016rgu,Damour:2017ced}.
The observation of gravitational waves from inspiraling black holes by the LIGO interferometer \cite{Abbott:2016blz}, evidently,
strongly motivates precision studies of the dynamics of these systems.
In this paper we will analyze a distinct question:
can we find analytically tractable models, in a theory which is realistic or not, of an orbiting pair of black holes.

Analytically tractable models generally possess additional symmetries and conservation laws.
This is famously the case in the Newtonian limit, where
the fact that we get closed Keplerian elliptical orbits is a consequence of the conservation of the
Laplace-Runge-Lenz vector, which points in the direction of the eccentricity.
In Einstein gravity, this vector is no longer conserved: observable effects include the precession of Mercury's perihelion angle.
The analytic model we are looking for will therefore differ from Einstein's theory in such a way that it will
not be a precise theory of nature, but hopefully it can remain a useful model.

The most optimistic outcome would be to find a system that is completely integrable.
Integrable field theories in higher dimensions are few and far between -- the only presently known example of an interacting
four-dimensional quantum field theory whose spectrum can be computed exactly is $\mathcal{N}=4$ super Yang-Mills in the planar limit \cite{Beisert:2010jr,Gromov:2014caa}, and deformations of it \cite{Gurdogan:2015csr}.
In fact quarkonium-like bound states in this theory are known to enjoy a Laplace-Runge-Lenz-like symmetry
which survives relativistic corrections \cite{Caron-Huot:2014gia}.
It would clearly be interesting if there existed a relativistic quantum theory of gravity that is integrable in the same sense;
this would provide a testbed for perturbative and nonperturbative methods but could also shed light on various questions in quantum gravity.
While complete integrability might be too much to ask for
(any gravitational theory will contain the three-body problem),
clearly any amount of additional symmetries would make a model valuable.

The best candidate for a such a symmetrical model of gravitational dynamics is $\mathcal{N}=8$ supergravity
\cite{Cremmer:1978ds,Cremmer:1979up}.
In addition to being closely tied to $\mathcal{N}=4$ super Yang-Mills by squaring relations \cite{Bern:2010ue,Bern:2008qj,Bern2002},
there have been some arguments that $\mathcal{N}=8$ should be the simplest quantum field theory, see \cite{ArkaniHamed:2008gz}.
Despite many interesting patterns and an overall feeling of simplicity,
observed for example in studies of the possible ultraviolet finiteness of the quantum theory \cite{Bern:2013uka,Bern:2018jmv},
an underlying symmetry principle has thus far not been identified.
In this paper, we propose to look at a simple subsystem:
We will investigate the symmetries of a pair of orbiting half-BPS black holes.

Systems of supersymmetric black holes have been extensively studied, in particular, supersymmetric bound states thereof.
For example it is well-known that if the charge vectors of two extremal holes are aligned, the electric repulsion exactly cancels
the gravitational (and scalar) attraction, leading to static solutions which experience vanishing force.
Static configurations which carry angular momentum but are overall BPS, involving magnetic and electric charges,
also exist at special positions where the forces cancel \cite{Denef:2002ru,Bates:2003vx,Anninos:2012gk,Avery:2007xf}. Partition functions in less supersymmetric cases, and the related wall-crossing phenomenon, have also been extensively studied.
We will be interested in pairs which are \emph{not} overall BPS and dynamically rotate around each other;
for generic charge vectors that are not aligned, the force then does not cancel and one generically finds
Kepler-like elliptical orbits.

With decreasing orbit radius, relativistic corrections become important.
These include effects which, in the case of General Relativity, lead to the precession of orbits as is the case of Mercury.
The precession rate in $\N$ supergravity is however expected to be different, because
this theory contains additional massless particles and thus other long-range forces.
As a first step, in this paper we determine whether the precession is zero or not at first nontrivial order.
that is, whether a conserved Laplace-Runge-Lenz vector continues to exist.
This will be interpreted as a diagnosis of whether $\mathcal{N}=8$ indeed possess hidden symmetries.
The question can be sharpened to make sense even including radiation effects, as discussed in conclusion.

We will work to first nontrivial order in two distinct (but overlapping) power-counting schemes:
the post-Newtonian expansion, which is in low velocity and large distances, and the probe limit $m_B\to 0$,
where one of the black holes is very light and follows geodesic motion in the background of the bigger black hole.
In the first case, following a standard strategy, we rely on the fact that closed and open (scattering) orbits
are controlled by the same effective Hamiltonian, which we extract from the scattering amplitudes of black holes.
This approach goes beyond the probe approximation since it remains applicable even for similar-mass black holes. A chief advantage for us is the robustness of this method: the symmetries of $\N$ can be straightforwardly implemented
at the level of on-shell scattering amplitudes, without requiring the supergravity Lagrangian nor subtle cancelations between terms;
we anticipate that this method will be particularly effective at higher orders.
For the second method, we will rely on the explicit realization of supersymmetric black holes as black branes in IIA supergravity,
and the corresponding Born-Infeld action.

This paper is organized as follows.
In section \ref{setup} we present our setup and parametrize the most general misalignment angle between
the charge vectors of two half-BPS black holes. We also introduce the relevant on-shell superspace.
In section \ref{amplitude} we compute the on-shell scattering amplitudes to one-loop order,
and show that they satisfy the no-triangle property;
subtleties arising in the presence of magnetic charges are pointed out.
In section \ref{precession} we deduce from this result the effective Hamiltonian at 1PN order.
Using a recent integrand-level subtraction scheme, we in fact obtain the effective Hamiltonian to first post-Minkowski order.
We show that the no-triangle property at this order is precisely equivalent to the absence of precession.
Finally, in section \ref{probe} we discuss the dynamics in the probe limit using a ten-dimension realization of the setup,
where D0,D2 and D6 branes with flux probe a D6 background.
We conclude in section \ref{sec:conclusion} with a conjecture and future outlook.
Five appendices give further detail on: the parametrization of charge angles; their realization in string theory;
a determination of the amplitudes without using superspace; further comments on monopole scattering amplitude and the calculation of quantum-mechanical matrix elements.


\section{Setup}
\label{setup}

We consider a system of two large half-BPS (extremal) black holes in $\N$ supergravity, with masses $m_A$ and $m_B$ each carrying electric and/or magnetic charges. As mentioned, we are interested in the case where the charge vectors
are not aligned, so that the force is nonvanishing in the Newtonian limit and the black holes dynamically orbit around each other.
In this section we describe how to parametrize the charges of two such black holes,
and how we will use on-shell superspace to simplify the scattering problem.

\subsection{Black hole charges}

There are 28 possible types of electric charges in this theory, with 28 corresponding magnetic charges.
These can be conveniently packaged into the real and imaginary part of an antisymmetric $8\times 8$ matrix $C^{IJ}$.
Specifically, in the scattering problem, the matrix arises as
the central charge of the $\mathcal{N}=8$ supersymmetry algebra when the black holes are studied in isolation:
\begin{equation}
 \label{susy algebra}
\{ q^I_\alpha, q_{J\dot\alpha} \} = \delta^I_J p_{\alpha\dot\alpha},\qquad
    \{q^I_{\alpha},q^J_{\beta}\}=2\epsilon_{\alpha\beta}C^{IJ},\qquad
   \{\bar{q}_{I\dot\alpha},\bar{q}_{J\dot\beta}\}=2\epsilon_{\dot\alpha\dot\beta}C_{IJ},
\end{equation}
where $I,J=1,...,8$ are $R$-symmetry indices, $\alpha,\dot\alpha=1,2$ spinor indices,
and $C_{IJ}=(C^{JI})^*$.
We work with mostly-plus metric and our sign conventions follow \cite{Elvang:2013cua}.
Central charges for half-BPS states satisfy an additional constraint that all their skew-eigenvalues
are equal to each other, or equivalently, the product $C C^\dagger$ is proportional to the identity matrix:
\begin{equation}
\label{C constraints}
C^{IJ}C_{JI'}= M^2  \delta^I_{I'}\qquad \arg \text{Pfaffian}(C)=0.
\end{equation}
The Pfaffian constraint encodes the statement that the charges are all in the same U-duality orbit \cite{Hull:1994ys}.
This will be seen as a consistency requirement on the amplitude in section \ref{amplitude},
and in section \ref{probe} to arise naturally from string constructions.

\def\C{\mathcal{C}}
In the full, nonperturbative, string theory, the charges $C$ are quantized and take values within a known charge lattice \cite{Hull:1994ys}.
Since in this paper we will be working in the classical limit where the charges are very large,
the quantization condition will play no role.
Any charge can then be obtained from another using the (continuous) SU(8) subgroup of the E${}^{7,7}$ duality symmetry which
preserves the asymptotic values of the scalar fields.
Indeed it is easy to show that the space of central charges satisfying eqs.~(\ref{C constraints}), let's call it $\C$,
is transitive under the action of SU(8).

Using a SU(8) rotation we can always put the first charge into the following canonical form:
\begin{equation}
\label{canonical form}
C_A=m_A\times
\begin{pmatrix}
0 & 1_{4\times 4} \\
-1_{4\times 4} & 0
\end{pmatrix}.
\end{equation}

As mentioned, it will be important for us that the charge vectors of the two black holes are \emph{not} aligned with each other.
Let us thus describe the possible configurations of two charges.
According to the definition of symplectic groups, $C_A$ is preserved by any SP(8,C) transformation:
\begin{equation}
    \Lambda^T C_A\Lambda=C_A, \qquad \Leftrightarrow \qquad \Lambda \in {\rm Sp(8,C)}.
\end{equation}
The stabilizer subgroup of the charge $C_A$ is therefore 
\begin{equation}
\label{eq:3}
    H={\rm Sp(8,C)}\cap {\rm SU(8)}, \qquad H^TC_AH=C_A
\end{equation}
Since SU(8) acts transitively, the space of possible (classical) charges is the coset $\C=$SU(8)/$H$.
The infinitesimal generators of $H$ can be described explicitly as follows:
\begin{equation}
   \mbox{generators of } H: \begin{pmatrix}
     A & B\\   B^\dagger & -A^T
    \end{pmatrix}
\end{equation}
where $B=B^T$ is a complex symmetrical $4\times 4$ matrix and $A=A^\dagger$ is an Hermitian matrix.
Thus $\dim H=36$, giving the dimension of the space of charges as
\begin{equation}
\label{eq:dim}
    \dim \C= \dim {\rm SU(8)}-\dim H = 63-36 = 27.
\end{equation}
This agrees with a direct analysis of the constraints (\ref{C constraints}) (which give 29
constraints on the 56 real parameters of an $8\times 8$ complex antisymmetric matrix).

We are interested in configurations of \emph{pairs} of charges,
and since $H$ does not change the charge of the first black hole, this means we can physically identify
charges which differ by $H$ action. The space of inequivalent charge pairs is therefore the
double coset $H\bs$SU(8)$/H$.

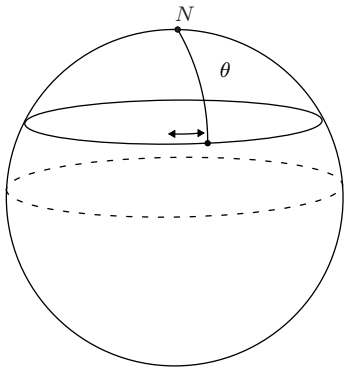
\begin{figure}[h]
    \centering
    \definecolor{sqsqsq}{rgb}{0.12549019607843137,0.12549019607843137,0.12549019607843137}
\begin{tikzpicture}[line cap=round,line join=round,>=triangle 45,x=1cm,y=1cm,scale=0.5]
\clip(-1.68,-6.09) rectangle (10.82,4.5);
\draw [line width=0.5pt] (5.62,-0.92) circle (4.476114386384691cm);
\draw [rotate around={-179.1835096827783:(5.613745415203001,-0.6412599556612641)},line width=0.5pt,dash pattern=on 2pt off 4pt] (5.613745415203001,-0.6412599556612641) ellipse (4.470115500845525cm and 0.7941236210437633cm);
\draw [rotate around={0.5866077925924609:(5.586278292584797,1.0908479555475026)},line width=0.5pt] (5.586278292584797,1.0908479555475026) ellipse (3.9529672848579245cm and 0.5857010569646879cm);
\draw [shift={(0.5506876482249646,0.580063953609407)},line width=0.5pt]  plot[domain=-0.008420778791939476:0.5240135882478231,variable=\t]({1*5.946427156292505*cos(\t r)+0*5.946427156292505*sin(\t r)},{0*5.946427156292505*cos(\t r)+1*5.946427156292505*sin(\t r)});
\draw [shift={(5.774511627906971,5.831255813953497)},line width=0.5pt]  plot[domain=4.672321349029668:4.814936278600248,variable=\t]({1*5.057919952464473*cos(\t r)+0*5.057919952464473*sin(\t r)},{0*5.057919952464473*cos(\t r)+1*5.057919952464473*sin(\t r)});\begin{scriptsize}
\draw [fill=sqsqsq] (5.699210857703213,3.555413460231571) circle (2pt);
\draw[color=sqsqsq] (5.8870697674418535,3.9853023255813933) node {$N$};
\draw [fill=black] (6.496903976629088,0.5299909977036213) circle (2pt);
\draw[color=black] (6.967627906976736,2.499534883720929) node {$\theta$};
\draw [fill=black,shift={(5.571906976744181,0.7773953488372091)},rotate=90] (0,0) ++(0 pt,3pt) -- ++(2.598076211353316pt,-4.5pt)--++(-5.196152422706632pt,0 pt) -- ++(2.598076211353316pt,4.5pt);
\draw [fill=black,shift={(6.292279069767436,0.7999069767441858)},rotate=270] (0,0) ++(0 pt,3pt) -- ++(2.598076211353316pt,-4.5pt)--++(-5.196152422706632pt,0 pt) -- ++(2.598076211353316pt,4.5pt);\end{scriptsize}\end{tikzpicture}
    \caption{$\theta$ is parametrizing $SO(2)\bs$SO(3)$/SO(2)$, {\it i.e.}, the angle between north pole and different SO(2) orbits }
    \label{fig:sphere}
\end{figure}

It is helpful to recall a familiar example:
pairs of points on a two-sphere modulo SO(3) rotations,
as depicted in fig.~\ref{fig:sphere}.
SO(3) acts transitively on the sphere so we can always rotate the first point to the North pole.
Its stabilizer is SO(2) (rotations around the $z$ axis). The second point is then at a general position on the sphere, which is the coset SO(3)/SO(2).
However, acting with SO(2) rotations on that second point produces rotationally equivalent pairs, so points along the double-sided arrow in fig.~\ref{fig:sphere}
are equivalent. The double coset SO(2)$\bs$SO(3)/SO(2) is one-dimensional and captures the invariant property of the pair:
the angle between the two points, denoted by $\theta$ in fig.~\ref{fig:sphere}.

The double coset $H\bs$SU(8)$/H$ similarly parametrizes possible angles between two half-BPS black hole charges.
A brief computation in appendix \ref{app:coset} shows that its dimension is 3,
so that 3 angles are necessary to specify the misalignment between their electric and magnetic charges.
In fact it is easy to write down a representative of this coset:
\begin{equation}
\label{second charge}
C_B=m_B\times
\begin{psmallmatrix}
 &  &  &  & e^{i\phi_1} &  &  &  \\
 &  &  &  &  & e^{i\phi_2} &  &  \\
 &  &  &  &  &  & e^{i\phi_3} &  \\
 &  &  &  &  &  &  & e^{i\phi_4} \\
-e^{i\phi_1} &  &  &  &  &  &  &  \\
 & -e^{i\phi_2} &  &  &  &  &  &  \\
 &  & -e^{i\phi_3} &  &  &  &  &  \\
 &  &  & -e^{i\phi_4} &  &  &  & \\
\end{psmallmatrix}
,\qquad \prod_{i=1}^4 e^{i\phi_i}=1.
\end{equation}
At the infinitesimal level it is easy to prove that any pair of charges can be put into this form.
We will thus study two black holes with the charges (\ref{canonical form}) and (\ref{second charge}). We will refer to this scattering with three free angles as the \emph{three angle case}.
\subsubsection{Magnetic charge and special cases without it}

The real and imaginary part of $C_B$ represent respectively its electric and magnetic charges.
Therefore, for generic angles $\phi_I$, we are scattering dyons. A useful way to quantify the magnetic charges is the Dirac bracket between
the charges of the two black holes, defined as the product of their 28 electric and magnetic charges:
\be
 \langle C_A,C_B \rangle \equiv \sum_{i=1}^{28} 
 \big(q^i_A g^i_B-g^i_A q^i_B\big)= 2\pi G_N{\rm Im}\,{\rm Tr}\big[C_A C_B^*\big].
\ee
The Dirac quantization condition requires that this should be $2\pi$ times an integer $N_{AB}$; for the above configurations this evaluates to
\be
\label{NABdef}
\frac{1}{2\pi} \langle C_A, C_B \rangle = 2G_Nm_Am_B\sum_{I=1}^4 \sin(\phi_I) \equiv N_{AB}\,.
\ee
As we will see in the next section,
the scattering of a pair of objects with $N_{AB}\neq 0$ is technically more involved because in this case
the S-matrix does \emph{not} depend only on dot products of external momenta.
For this reason it will be technically useful to also consider special cases where there are no magnetic charges.

It is easy to show that the condition $N_{AB}=0$ implies that the angles organize into two pairs:
$\phi_3=-\phi_1$, $\phi_4=-\phi_2$.
We'll thus refer to this case as the \emph{two angle case}.
Then time-reversal symmetry is preserved,
and in fact, up to a similarly transformation $C_B\to U C_B U^T$ with $U\in H$,
we can assume that $C_B$ is real so all charges are electric:
\be \label{two-angles}
 C_B\Big|_{\rm two-angles} = m_B
\begin{pmatrix} 0 & \tilde{C}_B \\ -\tilde{C}_B^T &0 \end{pmatrix},
\qquad
\tilde{C}_B =  \begin{pmatrix}
 \cos(\phi_1) & \sin(\phi_1) & 0 & 0 \\
-\sin(\phi_1) & \cos(\phi_1) & 0 & 0 \\
0 & 0 & \cos(\phi_2) & \sin(\phi_2) \\
 0 & 0 & -\sin(\phi_2) & \cos(\phi_2) \end{pmatrix}.
\ee
Finally, let us comment on a one-angle case where $\phi_1=\phi_2=-\phi_3=-\phi_4$.
This case can be realized in terms of higher-dimensinonal graviton scattering if one views $\mathcal{N}=8$ supergravity
as 10-dimensional type-IIA supergravity compactified on $T^6$.
The magnitude of the momenta
along the $T^6$ then provide the masses $m_A, m_B$, while their relative orientation
provides the one angle (this can be easily seen from the expressions in appendix \ref{app: charges}).
The integrand in this case has been extensively studied and is currently known even to five loops,
thanks to a recent calculation motivated by the study of ultraviolet divergences
\cite{Green:1982sw,Bern:1998ug,Bern:2008pv,Bern:2013uka,Bern:2018jmv}.

\subsection{On-shell superspace}
\label{sec:superspace}

The force between the black holes
originates from the exchanged gravitons and their superpartners (graviphotons and scalars).
Since they are in the same multiplet, the relative strength of their couplings are fixed by supersymmetry.
This information is conveniently packaged into a superamplitude.
This will prove especially useful for computing the 
post-Newtonian corrections arising from exchanging a pair of (super)gravitons.

An important point is that, because we are interested in long-range interactions,
it suffices to consider exchanged gravitons that are on-shell, that is their four-momenta are null.
Mathematically, this is simply because the Fourier transform relates the long-distance forces to discontinuities with respect to the momentum transfer,
which can be computed (see section \ref{discontinuity}) in terms of cut diagrams thanks to the Cutkowski rules.
Furthermore, for extracting the force from the scattering problem rather than directly studying bound orbits,
it suffices to analyze the transformation properties of asymptotic states. The action of supersymmetry then simplifies dramatically.

For a massless graviton of four-momentum $p$, a standard construction packages
the $2^8$ states of its multiplet into a polynomial in $8$ Grassmann variables $\eta^I$, see \cite{ArkaniHamed:2008gz,Elvang:2013cua}.
We take the component with $\eta=0$ to be the positive-helicity graviton,
with states of lower helicity obtained by adding more $\eta$'s.
The algebra then acts on the one-particle state $\ket{\eta}$ as
\begin{equation}
\begin{split}
[q^I| \equiv &[p|\frac{\partial}{\partial\eta_I} \qquad I={1,2....8}\\
|\bar{q}_I\rangle \equiv &|p\rangle\eta_I \qquad I={1,2....8}
\end{split} \label{susy massless}
\end{equation}
where $[p|$ and $|p\rangle$ are two-spinors defined via
the spinor-helicity factorization $p = |p\r\l p|$ on the on-shell graviton momentum.
It can be easily checked that the commutation relations (\ref{susy algebra})
of supercharges are satisfied.

Massive half-BPS black holes again form multiplets with $2^8$ states,
which we can again span using 8 Grassmann variables.
Because the rest frame of the black hole now leaves an unbroken SU(2) (little group) rotational symmetry, we cannot parametrize the 16 $q^I_\alpha$ without breaking either of the rotational or SU(8) symmetry.
It will be natural here to preserve the rotational symmetry --following \cite{Arkani-Hamed:2017jhn}-- but break the SU(8) $R$-symmetry,
as the latter is already broken by the central charges of the black holes.

To find how to parametrize the algebra, we note from the susy algebra \eqref{susy algebra} and charge (\ref{canonical form})
that the $q_I$'s, acting on the first black hole, do not commute:
\begin{equation}
\begin{split}
\{q^I_{\alpha},q^{J+4}_{\beta}\}&=2\epsilon_{\alpha\beta} m_A \delta^{IJ}.
\end{split}
\end{equation}
This suggests representing the first four $q^I_\alpha$ as Grassmann derivatives,
$\partial/\partial \theta_{\alpha,I}$ with $I=1,2,3,4$,
acting on a collection of four two-spinors $[\theta_I|$, and then the remaining $q^I$ by multiplication operators.
From the commutation relations,
$\bar{q}_{I\dot\alpha}$ and $\bar{q}_{I+4\,\dot\alpha}$ must then become multiplication and derivative operators, respectively,
leading to the following parametrization:
\begin{equation}
\label{susy massive}
\begin{aligned}
~[q^I |&=\sqrt{m_A}[\frac{\partial}{\partial \theta_I}|, &\qquad
[q^{I+4}|&=\sqrt{m_A} [\theta_I |, &\qquad I&={1,2,3,4},
\\
|\bar{q}_{I}\rangle&= \frac{1}{\sqrt{m_A}} |p_A|\theta_I],  &\qquad
|\bar{q}_{I+4}\rangle&= \frac{1}{\sqrt{m_A}} |p_A|\frac{\partial}{\theta_I}], &\qquad I&={1,2,3,4}.
\end{aligned}
\end{equation}
These acts on single particle states of the first black hole,
and satisfy all the commutation relation simultaneously.

For the second black hole, whose charge in eq.~(\ref{second charge})
differs by some phases, we simply multiply the first four components as
$q^I \mapsto e^{i\phi_I}q^I$ and $\bar{q}_I \mapsto \bar{q}_I e^{-i\phi_I}$ for $I=1,2,3,4$, but leave the generators unchanged for $I=5,6,7,8$.


\section{Tree and one-loop scattering amplitudes}
\label{amplitude}
\subsection{Black hole superamplitude and distinguished component}

\begin{figure}
\centering
 \scalebox{0.75}{
\begin{feynman}
    \fermion[label=$p_B$,labelDistance=0.25]{2.60, 0.20}{1.80, 1.00}
    \fermion[label=$p_A^{\prime}$, labelDistance=0.30, labelLocation=0.75]{1.00, 1.80}{0.20, 2.60}
    \fermion[label=$p_B^{\prime}$]{1.80, 1.80}{2.60, 2.60}
    \fermion[label=$p_A$,labelDistance=0.35]{0.20, 0.20}{1.00, 1.00}
    \parton{1.40,1.40}{0.50}\end{feynman}}
\label{fig:kinematics}
\caption{Kinematics of our $2\to 2$ scattering of black holes. Time runs upward.}
\end{figure}
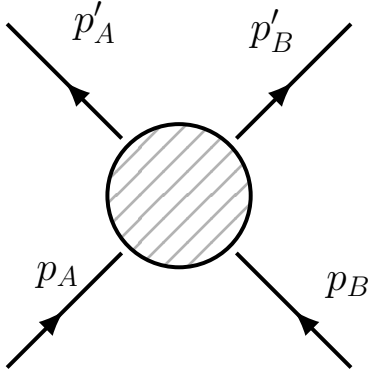

The scattering amplitudes between the various $2^8$ internal states of our black holes are not independent,
but turn out to be proportional to a single function of the Mandelstam invariants.
These relations will greatly help our computation and follow from the Ward identities (see ~\cite{Elvang:2013cua}) stating that the supercharges annihilate the superamplitude:
\begin{equation}
\label{ward1}
Q^I \mathcal{A}_n=0 \qquad\mbox{and} \qquad \bar{Q}_I\mathcal{A}_n=0,
\end{equation}
where $Q^I$ is the sum over the supercharges of incoming and outgoing particles:
\begin{equation}
\label{ward2}
Q^I=\sum_{i=1}^n q^I_i \qquad \mbox{and} \qquad \bar{Q}_I=\sum_{i=1}^n \bar{q}_{Ii}, \qquad I={1\ldots 8}.
\end{equation}
Using the generator (\ref{susy massive}) for the massive particles,
for the $2\to 2$ scattering of black holes we find that 16 of these constraints are simple multiplicative operators:
\begin{equation}
\begin{aligned}
\label{eq:charges}
[Q^I|=&\sqrt{m_A} [\theta_{1,I-4}|+\sqrt{m_A} [\theta_{2,I-4}|+\sqrt{m_B} [\theta_{3,I-4}|+\sqrt{m_B} [\theta_{4,I-4}|\,,\qquad I={5,6,7,8},\\
~|\bar{Q}_I\r=&\frac{1}{\sqrt{m_A}} |p_A|\theta_{1,I}]+\frac{1}{\sqrt{m_A}} |p_A^{\prime}|\theta_{2,I}]+\frac{e^{-i\phi_I}}{\sqrt{m_B}} |p_B|\theta_{3,I}]+\frac{e^{-i\phi_I}}{\sqrt{m_B}} |p^{\prime}_B|\theta_{4,I}]\,,\qquad I={1,2,3,4}.
\end{aligned}
\end{equation}
This implies that the $2\to 2$ amplitude is proportional to 16 Grassmann $\delta$-functions
\begin{equation}
\label{eq:amp}
\begin{split}
\mathcal{A}_{\Phi\Phi\Phi\Phi}=& \frac{A}{(m_Am_B)^4}\times
\prod_{I=1}^4 \delta^2(\bar{Q}_I) \prod_{I=5}^8 \delta^2(Q^{I})
\end{split}
\end{equation}
which is analogous to the familiar $\delta^{16}(Q)$ in the massless case (see \cite{ArkaniHamed:2008gz}).
The remaining Ward identities then take the form of derivative operators, for example
\be
Q^I=  \sqrt{m_A}[\frac{\partial}{\partial \theta_{1,I}}|- \sqrt{m_A}[\frac{\partial}{\partial \theta_{2,I}}|+ \sqrt{m_B}e^{i\phi_I}[\frac{\partial}{\partial \theta_{3,I}}|- \sqrt{m_B}e^{i\phi_I}[\frac{\partial}{\partial \theta_{4,I}}|\qquad I={1,2,3,4}
\ee
This simply imposes that $A$ in eq.~(\ref{eq:amp}) is independent of Grassmann variables.
The upshot is that the amplitude for arbitrary internal states of the black holes are all proportional to the same function of (bosonic) momenta,
$A$.

To match with an effective Hamiltonian in the next section, it will be convenient to restrict our
attention to a particularly simple, distinguished state in the multiplet, for which fermion exchange diagrams vanish.
This is easily done by maximizing some SU(8) charges, for example by choosing initial internal states
that are identical complex scalars $\phi$ at the bottom of the multiplet, $\theta_A=\theta_B=0$.
The scattering amplitude for this component is then 
obtained by extracting the term with the maximum power of $\theta'_A$, $\theta'_B$,
that is the coefficient of $[\theta'_A \theta'_A]^4[\theta_B' \theta_B']^4$, and vanishing powers of $\theta_A,\theta_B$.
A short computation from eq.~(\ref{eq:amp}) gives
\be
\mathcal{A}_{\phi\phib\phi\phib} = A \times  \prod_{I=1}^4 \left[ \cosh\eta_{AB}-\cos\phi_I\right] \label{eq:amp phi}
\ee
where $\cosh\eta_{AB}\equiv \frac{-p_A{\cdot}p_B}{m_Am_B}$, defining $\eta_{AB}$ as the relative rapidity between the incoming black holes.
Note that we used $\prod e^{i\phi_I}=1$ to simplify, and that the product could be written
in terms of BPS thresholds: $\prod_i \big(s-|m_A+ie^{i\phi_I}m_B|^2\big)$.

When the charges are aligned and the velocity vanishes, the product in (\ref{eq:amp phi}) yields a suggestive $(0)^4$.
The fourth power will get reduced to second-order zero once we account for $A$ below,
but it is nonetheless largely responsible for the vanishing force between static same-charge black holes.

In the rest of this section we will compute the amplitude at tree and one-loop order. In fact, we are only interested in the long-distance
interactions, determined by the singular part as $t\to 0$, so we will focus on the $t$-channel cuts.
We will thus include the singularities from exchange of one and two (super)gravitons.

\subsection{Tree-level black hole scattering}

At tree-level (thus giving the Newtonian force at very large impact parameter)
the two black holes can exchange graviton, graviphotons and scalars (but not fermions, for the specific initial state just described).
The relative coefficients are, of course, determined by supersymmetry, and this calculation provides a useful warm-up exercise
before we go to loop level.
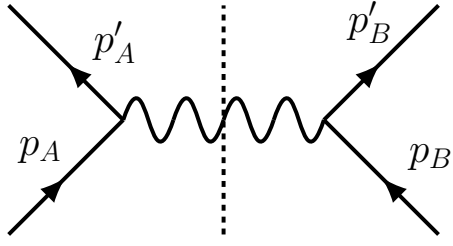
\begin{figure} 
\centering
\scalebox{0.75}{
\begin{feynman} 
    \fermion[label=$p_A^{\prime}$, labelDistance=0.2]{1.00, 1.00}{0.20, 1.80}
    \fermion[label=$p_{A}$, labelDistance=0.40, labelLocation=0.38]{0.20, 0.20}{1.00, 1.00}
    \electroweak{1.00, 1.00}{2.40, 1.00}
    \dashed[showArrow=false]{1.70, 0.20}{1.70, 1.80}
    \fermion[label=$p_{B}$, labelDistance=0.20]{3.20, 0.20}{2.40, 1.00}
    \fermion[label=$p_B^{\prime}$, labelDistance=0.40]{2.40, 1.00}{3.20, 1.80}
\end{feynman}}
\label{fig:1}
\caption{$2\to 2$ scattering of black holes, tree-level}
\end{figure}
Specifically, we first determine the 3-point vertices for emission of a supergraviton from a massive spinless black hole
(which determine the $t$-channel pole, illustrated in the diagram in fig.~\ref{fig:1}).
The Ward identities involve both the massive representation (\ref{susy massive}) and the massless one (\ref{susy massless}).
Again 16 of the supercharges are simple multiplication with respect to the $\theta$'s,
however we now have both multiplication and derivatives with respect to $\eta$:
\begin{equation}
\label{Ward 3}
\begin{split}
[Q^I|=&\sqrt{m_A} [\theta_{A,I-4}|+\sqrt{m_A} [\theta_{A',I-4}|+[3|\frac{\partial}{\partial\eta_{3,I}}, \qquad I={5,6,7,8}\\
|\bar{Q}_I\rangle=&\frac{1}{\sqrt{m_A}} |p_A|\theta_{A,I}]+\frac{1}{\sqrt{m_A}} |p_A^{\prime}|\theta_{A',I}]+|3\rangle\eta_{3,I}, \qquad I={1,2,3,4}.
\end{split}
\end{equation}
The solution is
\begin{equation}
\label{eq:c1}
\mathcal{A}^L_{\Phi\Phi\eta}=\frac{A^L_{\Phi\Phi\eta}}{m_A^2}
\times\prod_{I=1}^4 \delta^2(\bar{Q}_I)
\exp\left(\frac{\sum_{I=1}^4\Big(\sqrt{m_A}[\theta_{A,I}+\theta_{A',I},\xi]\Big)\eta_{I+4,3}}{[3\xi]}\right),
\end{equation}
and one finds that the other Ward identities simply impose that $A_{\Phi\Phi\eta}$ is independent
of Grassmann variables.
Extracting the $\phi\phib$ complex scalar amplitude by setting $\theta_A=0$ and using the $\delta$-function to eliminate
$\theta_{A'}$ from the exponent, we find simply
\begin{equation}
\label{lhs}
  \mathcal{A}^L_{\phi\phib\eta}=A^L_{\Phi\Phi\eta}\exp\left(x_A\sum_{I=1}^4\eta_I\eta_{I+4}\right), \qquad
  x_A\equiv \frac{\langle 3|p_A|\xi]}{[3\xi]m_A} = \left(\frac{[3|p_A|\xi\rangle}{\langle 3\xi \rangle m_A}\right)^{-1}.
\end{equation}
The quantity $x_A$ is defined when all the momenta are on-shell and is independent
of the reference $\xi$, see \cite{Guevara:2017csg,Arkani-Hamed:2017jhn}.

To find the overall normalization one may simply extract the $g_{++}$ helicity component by setting $\eta=0$ and compare with
the standard three-point vertex in General Relativity, which is the matrix element of the basic interaction Lagrangian:
\begin{equation}
    \mathcal{L}_{\rm int}=\sqrt{8\pi G_N}h_{\mu\nu}T^{\mu\nu},
\end{equation}
where $g_{\mu\nu}\equiv \eta_{\mu\nu}+\kappa h_{\mu\nu}$ is the metric tensor (with $\kappa=\sqrt{32\pi G_N}$ the gravitational coupling),
and $T^{\mu\nu}$ is the energy-momentum. The matrix element between scalars then gives:
\begin{equation}
\label{eq1}
    \langle p_A^{\prime}|T^{\mu\nu}(x=0)|p_A\rangle=\tfrac12 (p_A+p^{\prime}_{A})_{\mu}(p_A+p^{\prime}_{A})_{\nu}+O(q)
\end{equation}
Dotting into the polarization $g_{++}=\epsilon_+\epsilon_+$ with $\epsilon_+ = \sqrt{2}\frac{|3]\l \xi|}{\l 3\xi\r}$
thus gives the General Relativity prediction:
\begin{equation}
\label{A3g}
   \mathcal{A}^L_{\phi\phib g_{++}}=\sqrt{8\pi G_N}\frac{m_A^2}{x_A^2}.
\end{equation}
Thus the superamplitude is 
\be
 \mathcal{A}^L_{\phi\phib \eta} =\sqrt{8\pi G_N}\frac{m_A^2}{x_A^2}\exp\left(x_A\sum_{I=1}^4\eta_I\eta_{I+4}\right).
 \label{A3 L}
\ee
For the other side of the cut the result is modified simply by the phases in eq.~(\ref{lhs}),
\begin{equation}
 \label{A3 R}
 \mathcal{A}^R_{\phi\phib \eta} =\sqrt{8\pi G_N}\frac{m_B^2}{x_B^2}\exp\left(-x_B\sum_{I=1}^4e^{i\phi_I}\eta_I\eta_{I+4}\right),
\end{equation}
where the minus sign comes from the opposite sign of $p_3$.

By taking different powers of $\eta$ of eqs.~(\ref{A3 L}) and (\ref{A3 R}) one obtains the graviton, vector and scalar couplings, respectively. We notice that the exponent in all cases can be written in a SU(8) covariant form: $\frac12 C^{IJ}\eta_I\eta_J$, which shows that the electric-magnetic charge of the black holes ({\it i.e.}, their coupling to vector particles) is indeed given by $C^{IJ}$, as expected. This can be seen directly from the susy algebra, see appendix \ref{app: amplitudes}.

As a simple consistency check one may extract $\mathcal{A}_{\phi\phib g^{--}}$ from the superamplitudes and check that does give the complex conjugate result. We stress that, were it not for the Pfaffian constraint $\prod_I e^{i\phi_I}=1$,
the relation between the two helicity of the gravitons could not be satisfied. As far as we understand, such a relative phase between the two graviton helicity would then signal a pathology of the theory.

Now we proceed to multiply together the left hand side and the right hand side of the diagram in fig.~\ref{fig:1} (\eqref{A3 L} and \eqref{A3 R}), and integrate over $\eta$. The integral simply gives:
\be \label{tree eta integral}
 \int d^8\eta \exp\left(x_A\sum_{I=1}^4\eta_I\eta_{I+4}\right)
 \exp\left(-x_B\sum_{I=1}^4e^{i\phi_I}\eta_I\eta_{I+4}\right) = \prod_{I=1}^4(x_A-x_Be^{i\phi_I}).
\ee
Using that $\frac{x_A}{x_B}=e^{\pm \eta_{AB}}$ on the on-shell solutions, which is simply the boost factor,
the $t$-channel pole of the amplitude is thus
\begin{equation}
 \label{tree pole}
 \begin{split}
    \mathcal{A}^{\text{tree}}_{\phi\phib\phi\phib}\Big|_{\rm pole}=&
   8\pi G_N\frac{m_A^2m_B^2}{-t} \prod_{I=1}^4 2\sinh\left(\frac{\pm \eta_{AB}-i\phi_I}{2}\right)\,.
   \end{split}
\end{equation}
The sign $\pm\eta_{AB}$ differs for the two on-shell solutions and this is a symptom that in the context of
monopole scattering, the amplitude is not gauge invariant. This will be discussed in appendix \ref{app:monopole}. We will also further study magnetic charges in a classical context in section \ref{probe}.
For the moment, let us simply note that in the two-angle case, where $\phi_3=-\phi_1$ and $\phi_4=-\phi_2$
and there are no magnetic charges, the sign ambiguity disappears and
the amplitude is expressed in terms of $\cosh\eta_{AB}$ only (equivalently,
in terms of Mandelstam invariants only):\footnote{
Away from the $t=0$ pole, this could arise for example from a crossing-symmetric function $A$ in eq.~(\ref{eq:amp phi}):
\be
A=\frac{128\pi G_N m_A^4m_B^4}{t\big(s-|m_A+e^{i\phi_1}m_B|^2\big)\big(u- |m_A-e^{i\phi_2}m_B|^2\big)},
\ee
or the same thing with $\phi_1$ and $\phi_2$ exchanged.  Fixing it fully would require a more detailed analysis of the dynamics
but will not be required in this paper.}
\begin{equation} \label{result Atree 2-angles}
    \mathcal{A}^{\text{tree}}_{\phi\phib\phi\phib}\Big|_{\rm pole,\ 2-angles}=
   16\pi G_N\frac{m_A^2m_B^2}{-t}\times 2(\cosh\eta_{AB}-\cos \phi_1)(\cosh\eta_{AB}-\cos \phi_2).
\end{equation}
Now including the normalizing factor we define $\tilde{\AA}^{\rm tree}_{\phi\phib\phi\phib}=\frac{\AA^{\rm tree}_{\phi\phib\phi\phib}}{4E_AE_B}$. Keeping terms only up to the order $p^2$ we can write the velocity expansion of the tree-level amplitude as:
\begin{equation} 
\label{a00}
\begin{split}
  \tilde{\mathcal{A}}^{\text{tree}}_{\phi\phib\phi\phib}\Big|_{\rm pole,\ 2-angles}=&\frac{8\pi G_Nm_Am_B}{-t}(1-\cos\phi_1)(1-\cos\phi_2)
  \\&\left(1+\left(\frac{1}{2}\frac{1-\cos\phi_1\cos\phi_2}{(1-\cos\phi_1)(1-\cos\phi_2)}+\frac{m_Am_B}{(m_A+m_B)^2}\right)\frac{p^2}{\mu^2}+O(p^4)\right)
\end{split}
\end{equation}
where $\mu=\frac{m_Am_B}{(m_A+m_B)}$ is the reduced mass.

In the static limit $\eta_{AB}\to 0$, the interaction thus vanishes when the charges are aligned,
as expected from the vanishing force between such a BPS pair (but the degree of vanishing is lower than what was suggested by eq.~(\ref{eq:amp phi})).
\subsection{One-loop black hole scattering}
\label{oneloopbhscattering}
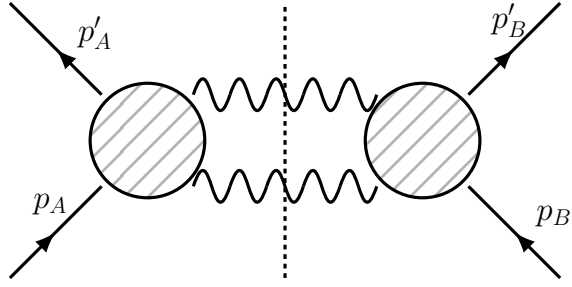
\begin{figure}[t]
    \centering
    \scalebox{0.6}{
    \begin{feynman} 
    \fermion[label=$p_A^{\prime}$,  labelDistance=0.2]{1.00, 1.80}{0.20, 2.60}
    \fermion[label=$p_B^{\prime}$,, labelDistance=0.35]{4.20, 1.80}{5.00, 2.60}
    \fermion[label=$p_{B}$,  labelDistance=0.2]{5.00, 0.20}{4.20, 1.00}
    \dashed[showArrow=false]{2.60, 0.20}{2.60, 2.60}
    \electroweak[]{1.80, 1.80}{3.40, 1.80}
    \electroweak[]{1.80, 1.00}{3.40, 1.00}
    \fermion[label=$p_{A}$, labelDistance=0.35]{0.20, 0.20}{1.00, 1.00}
    \parton{1.40,1.40}{0.50}
    \parton{3.80,1.40}{0.50}\end{feynman}
    }
    \caption{$2\to 2$ scattering of black holes, 1-Loop level}
    \label{fig:2}
\end{figure}

We are now ready to calculate the $t$-channel cut of the one-loop amplitude;
we again multiply tree diagrams, this time involving two (super)gravitons.
The helicity dependence is again fully determined by supersymmetry.
The Ward identities are as in eq.~(\ref{Ward 3}) but with one additional massless graviton $p_4$,
and because we still only have 4 external states we find again a unique solution:
\begin{equation}
\mathcal{A}^L_{\Phi\Phi\eta\eta}=
\frac{A^L_{\Phi\Phi\eta\eta}}{m_A^2}
\times \prod_{I=1}^4 \delta^2(\bar{Q}_I)
\exp\left(
\frac{\sqrt{m_A}}{[34]}
\sum_{I=1}^4\Big([\theta_{A,I}+[\theta_{A',I}\Big)\Big(|3]\eta_{4,I+4}-|4]\eta_{3,I+4}\Big)\right)
\end{equation}
where the prefactor is independent of Grassmann variables.
Setting $\theta_1=0$ and using the $\delta$-functions to eliminate $\theta_2$ from the exponent,
the amplitude for our complex scalars $\phi\phib$ is
\begin{equation} \label{AL phi phi eta eta}
\mathcal{A}^L_{\phi\phib\eta\eta}=
A^L_{\Phi\Phi\eta\eta} \exp\left(\frac{1}{[34]m_A}\sum_{I=1}^4
\Big(\langle q_{3,I}+q_{4,I}| p_A\Big)\Big(|3]\eta_{4,I+4}-|4]\eta_{3,I+4}\Big)\right)
\end{equation}
where we recall that $\langle q_{3,I}|= \langle 3|\eta_{3,I}$. This expression also admits a SU(8) covariant form which is given in eq.~\eqref{covariant A 2 etas}. The prefactor $A^L_{\Phi\Phi \eta\eta}$ can be fixed easily from
the $g_{++}g_{++}$ positive-helicity gravitons amplitude, obtained by setting $\eta_i=0$.
Requiring that this amplitude has the correct little-group weights and mass dimension, and
$s$- and $u-$ channel poles, one uniquely finds up to a constant factor:
\be
A^{L}_{\Phi\Phi \eta\eta}\propto
8\pi G_N \frac{m_A^4[34]^4}{(p_A\cdot p_3)(p_A\cdot p_4)(p_3\cdot p_4)}.
\ee
The constant factor can be fixed by requiring it to match with the square of eq.~\eqref{A3g}: 
\be
A^{L}_{\Phi\Phi\eta\eta}=8\pi G_N \frac{m_A^4[34]^4}{(s-m_A^2)t(u-m_A^2)}.
\ee
This agrees with the positive-helicity gravitons amplitude in General Relativity (see for example \cite{Holstein:2016fxh}).
The amplitude on the right is similar but has an extra phase $e^{i\phi_I}$ for each term
in the exponent as in eq.~(\ref{AL phi phi eta eta}).
Multiplying them, the cut integral is given as
\def\cut{\rm on-shell}
\be
 \AA^{\rm 1-loop}_{\phi\phib\phi\phib} \Big|_{\rm cut} \equiv
 \frac1{2!}\int \frac{d^4\ell}{(2\pi)^4} \left[\frac{1}{\ell^2(p_A-p^{\prime}_{A}-\ell)^2}\right]_{\cut}
 A^{L}_{\Phi\Phi\eta\eta} A^{R}_{\Phi\Phi\eta\eta} \times I
\ee
where $\left[ \frac{1}{X^2Y^2}\right]_{\cut} \equiv 2\pi \delta(X^2) 2\pi \delta(Y^2)\theta(X^0)\theta(Y^0)$ and
$I$ is the Grassmann integral:
\be
I= \int d^8\eta_3 d^8\eta_4
\exp\left(\frac{1}{[34]}\sum
\langle q_{3,I}+q_{4,I}|\left(\frac{p_A}{m_A} - e^{i\phi_I}\frac{p_B}{m_B}\right)\big(|3]\eta_{4,I+4}-|4]\eta_{3,I+4}\big)\right).
\ee
The integral nicely antisymmetrizes the $\lambda_3$ and $\lambda_4$ spinors,
which converts all spinor products to squares of four-vectors, leaving a simple result:
\be
I = \left( \frac{\langle 34\rangle}{[34]}\right)^4  \prod_{I=1}^4  \left(\frac{p_A}{m_A} -e^{i \phi_I}\frac{p_B}{m_B}\right)^2
 = 16\times\left( \frac{\langle 34\rangle}{[34]}\right)^4 \prod_{I=1}^4 (\cosh\eta_{AB} - \cos\phi_I).
\ee
Thus
\be
\begin{aligned}
\label{cutintegral}
\AA^{\rm 1-loop}_{\phi\phib\phi\phib} \Big|_{\rm cut} &=\frac{1}{2!} (8\pi G_N)^24(m_Am_B)^4 \prod_{I=1}^4\big(\cosh\eta_{AB} - \cos\phi_I\big)
\\&\qquad \times
 \int  \frac{d^4\ell}{(2\pi)^4} \left[\frac{1}{\ell^2\ell'^2}\right]_{\cut}
 \Bigg(\frac{\ell\cdot \ell'}{(p_A\cdot \ell)(p_A\cdot \ell')}\Bigg)\Bigg(\frac{\ell\cdot \ell'}{(p_B\cdot \ell)(p_B\cdot \ell^{\prime})}
 \Bigg)
\end{aligned}
\ee
where $\ell'\equiv p_A-p^{\prime}_{A}-\ell$.
This expression is understood to be valid only on the $t$-channel cut, that is when the two propagators
in the square bracket are set on-shell.
Using that $\ell\cdot \ell' = \ell\cdot (p_A^{\prime}-p_A)=\ell\cdot (p_B^{\prime}-p_B)$ on-shell,
the last line can also be written as the sum of $s$- and $u$-channel scalar box integrals:
\be \label{box integrals}
\int  \frac{d^4\ell}{(2\pi)^4}\left[\frac{1}{\ell^2\ell'^2}\right]_{\cut} \left(
  \frac{2}{(p_A\cdot \ell)(p_B\cdot \ell)}-  \frac{2}{(p_A\cdot \ell)(p'_B\cdot \ell)}\right).
\ee
This is the main result of this section. The essential conclusion is that the no-triangle property, previously found for massless graviton scattering (see \cite{BjerrumBohr:2006yw}), remains valid for massive case as well.
Although the above formulas are just for the $t$-channel cut, it is straightforward to see that the s-channel cut also gives only boxes,
so the full integrand (even away from the cut) is indeed the sum of two scalar box integrals. 

\subsection{Integrating the discontinuity}
\label{discontinuity}
Having recognized the discontinuity as that of a box integral, in principle at this stage we could use standardized results
for the box integral (see \cite{Denner:1991qq}). However, since we are only interested in the long-range force,
it is instructive to integrate directly the cut following a method by Feinberg and Sucher \cite{Feinberg:1988yw}
which manifests the relation between the discontinuity and the long-range potential.

The basic subtlety is that, in the $2\to 2$ kinematics that we are interested in,
$t<0 $ is purely spacelike for real external momenta. To reach the discontinuity we need to consider
complex external momenta. Feinberg and Sucher realized that this is actually a natural thing,
since, through the Fourier transform, large real distances are dominated by imaginary momenta.
Since $t$ is then a timelike invariant, it is useful to parametrize the external momenta
so that the momentum transfer is in the time direction.
This is related to the usual scattering kinematics by a double Wick rotation where one exchanges the usual time
and $z$ coordinates. As a result, in this frame the external momenta have purely imaginary vector parts, $\vec{r}$ and $\vec{r^{\prime}}$:
        \begin{equation}
            \vec{r}=i\xi_A\hat{r}, \qquad \vec{r^{\prime}}=i \xi_B\hat{r^{\prime}}
        \end{equation}
where the norms satisfy $\xi_A = m_A \sqrt{1-t/(4m_A^2)}$ and similarly for $\xi_B$.
The advantage of this frame is that the loop variables (the on-shell momenta $\ell,\ell'$ of the massless gravitons)
are real:
      \begin{equation}
      \begin{split}
      &\ell=(\frac{\sqrt{t}}{2},\vec{\ell}), \qquad  \ell^{\prime}=(\frac{\sqrt{t}}{2},-\vec{\ell}),
        \\
          &p_A=(\frac{\sqrt{t}}{2},\vec{r}), \qquad p'_A=(\frac{-\sqrt{t}}{2},\vec{r}),
        \\
          &p_B=(\frac{\sqrt{t}}{2},\vec{r^{\prime}}), \qquad p'_B=(\frac{-\sqrt{t}}{2},\vec{r^{\prime}}).
                   \end{split}.
      \end{equation}
This parametrization is valid for $t$ positive and small enough.
Since the norm of each vector is fixed, the integration depends only on angles
\begin{equation}
y =\hat{r}\cdot\hat{r}^{\prime}, \qquad w_A=\hat{\ell}\cdot \hat{r}, \qquad w_B=\hat{\ell}\cdot \hat{r}^{\prime}.
\end{equation}
Now to compute Disc${}_t \AA=2i\text{Im}\AA$, one can evaluate the phase space integral over the two on-shell propagators:
\begin{equation}
\begin{split}
 \int \frac{d^4\ell}{(2\pi)^4} \left[\frac{1}{\ell^2\ell'^2}\right]_{\cut}
=\int\frac{d^3\ell}{(2\pi)^32\ell_0}\frac{d^3\ell^{\prime}}{(2\pi)^32\ell^{\prime}_0}(2\pi)^4\delta^4(p_A+p_B-\ell-\ell^{\prime})
=\frac{1}{8\pi}\int \frac{d\Omega}{4\pi}.
\end{split}
\end{equation}
Inserting the integrand we are thus left with the following angular integral:
\begin{equation}
\begin{split}
\text{Disc}_t\,\AA^{\rm 1-loop}_{\phi\phib\phi\phib} =&\frac{i}{2!}(8\pi G_N)^24(m_Am_B)^4 \prod_{I=1}^4\big(\cosh\eta_{AB} - \cos\phi_I\big)\\&\times \frac{1}{2\pi(m_A^2-\frac{t}{4})(m_B^2-\frac{t}{4})}\int\frac{d\Omega}{4\pi}\Bigg(\frac{1}{(w_A^2+\frac{t}{4m_A^2-t})}\Bigg)\Bigg(\frac{1}{(w_B^2+\frac{t}{4m_B^2-t})}\Bigg).
\end{split}
\end{equation}
 The result of the integral has been originally produced by Feinberg and Sucher in \cite{Feinberg:1988yw} and is quoted in \cite{Holstein:2016fxh}. Without bothering with the complexity of the full answer, we take the $t\rightarrow 0$ limit of it and pick the leading term:
 \begin{equation}
\text{Disc}_t\,\AA^{\rm 1-loop}_{\phi\phib\phi\phib} =\frac{i}{\pi}(8\pi G_N)^2(m_Am_B)^2 \prod_{I=1}^4\big(\cosh\eta_{AB} - \cos\phi_I\big)\left[\frac{2\pi m_Am_B}{t(1-y^2-i0)^{\frac{1}{2}}}+O(t^0)\right].
\end{equation}
To undo the discontinuity we can simply use that $2\pi i$ is the discontinuity of $\log(-t)$,
which gives us the amplitude, modulo $t$-regular terms:
\begin{equation}
\label{a1}
\begin{split}
\AA^{\rm 1-loop}_{\phi\phib\phi\phib}
=\frac{(8\pi G_N)^2(m_Am_B)^3}{\pi} \prod_{I=1}^4\big(\cosh\eta_{AB} - \cos\phi_I\big)\left[\frac{1}{(1-y^2-i0)^{\frac{1}{2}}}\frac{\log(-t)}{t}+ O(t^0)\right].
\end{split}
\end{equation}
In the next section we will use the low-velocity expansion of this result.
It is useful to give an expression for y in terms of Mandelstam variables:
 \begin{equation}
 y=\hat{r}\cdot\hat{r}^{\prime}=\frac{2s+t-2m_A^2-2m_B^2}{4\xi_A\xi_B} = \cosh(\eta_{AB}) + O(t).
 \end{equation}
The amplitude in the non-relativistic normalization as in eq.~(\ref{a00}),
which we will need in the next section, to first post-Newtonian order is then, to order $O(t^0,p^3)$:
 \begin{equation}
\label{a2}
\begin{split}
\tilde{\AA}^{\rm 1-loop}_{\phi\phib\phi\phib}=16\pi G_N^2(m_Am_B)^2 \prod_{I=1}^4\big(\cosh\eta_{AB} - \cos\phi_I\big)\left[\frac{\mu}{\sqrt{-p^2-i0}}\left(1-\frac{p^2}{2m_Am_B}\right)\frac{\log(-t)}{t}\right].
\end{split}
\end{equation}
where $p$ is the center of mass momentum, and we recall that $\cosh\eta_{AB}$ also depends on $p$.
To conclude, let us contrast the present results with scalar scattering in General Relativity.
At tree-level only the two helicity of the gravitons contribute, which are the terms
with zero and eight powers of $\eta$ in eq.~(\ref{tree eta integral}), giving
\be
  \mathcal{A}^{\text{tree}}_{GR}\Big|_{\rm pole}=
 16\pi G_N\frac{m_A^2m_B^2}{-t} \times \cosh(2\eta_{AB})\,.
\ee
At one-loop, there is now a nonvanishing triangle coefficient, and one finds \cite{Holstein:2016fxh,BjerrumBohr:2002kt}:
\be
\label{gramp}
\tilde{\mathcal{A}}^{\rm 1-loop}_{GR}
= \frac{4\pi G_N^2m_Am_B\mu}{\sqrt{-p^2-i0}}\frac{\log(-t)}{t} + \frac{6G_N^2\pi^2(m_A+m_B)m_Am_B}{\sqrt{-t}} + O(t^0,p).
\ee
We see that the key difference is the absence of the $1/\sqrt{-t}$ term
in the $\mathcal{N}=8$ result \eqref{a2}. This stems directly (and is easily shown to be equivalent,
see subsection \ref{ssec: integrand subtractions} below) to the absence of scalar triangle integral
in the standard scalar box-scalar-bubble decomposition.
As we will now see, the absence of this term also implies the vanishing of the perihelion precession angle at 1PN order.

\section{Effective Hamiltonian and the precession}
\label{precession}

We are now ready to discuss the bound state problem.
The link with the scattering amplitude is provided by an effective Hamiltonian, since scattering and bound orbits are both controlled
by the same effective Hamiltonian describing near-instantaneous long-range forces between slow particles.
This approach is familiar in many contexts, including QED and 
QCD heavy-quark bound states \cite{Brambilla:2004jw}; for reviews in the context of gravity see
\cite{Donoghue:1995cz,Burgess:2003jk,Porto:2016pyg}.
There are many examples for deriving the angle of precession from such effective Hamiltonians in the literature,
for example \cite{Iwasaki:1971vb} and \cite{Hiida:1972xs} (see \cite{Bernard:2016wrg} for a recent 4PN calculation).
In this section we will restrict our attention to two-angle charge misalignment,
where there is no magnetic charge and time-reversal symmetry restricts the allowed terms.
We will find a direct connection between the precession angle and the amplitude.

\subsection{Effective Hamiltonian for bound states}

\def\Hzero{H_{\rm 0PN}}
\def\Hone{H_{\rm 1PN}}

The starting point for the post-Newtonian expansion is of course the lowest order Hamiltonian:
\begin{equation}
 \Hzero =  \frac{p^2}{2\mu}-\frac{\alpha}{r} \label{H0PN}
\end{equation}
where $\mu=\frac{m_Am_B}{m_A+m_B}$ is the reduced mass of the two black holes,
and $\alpha$ is proportional to $G_Nm_Am_B$ but with an additional dependence on the charge misalignement,
to be given shortly.
As above, $p$ denotes the spatial momentum of the first black hole in the rest frame of the pair.

For bound states we treat each term in $\Hzero$ as being of the same order
(the virial theorem equates the following time averages: $ \langle \frac{p^2}{\mu}\rangle=\langle\frac{\alpha}{r}\rangle$).
The post-Newtonian expansion is an expansion in
two small dimensionless parameters, to be treated of the same order:
\begin{equation}
    \frac{p^2}{\mu^2}\sim \frac{\alpha}{\mu r}\ll 1.
\end{equation}
We work in units where $\hbar=c=1$.
(Focusing on the classical limit, one usually also assumes $\hbar/(pr)\ll 1$, which gives a hierarchy between
the momenta $p$ and momentum transfer $q=p-p'\sim \hbar/r$. Below we will also consider
quantum mechanical bound states so we will not make this assumption here.)

The first post-Newtonian Hamiltonian contains velocity-squared corrections as well as all possible terms of the same order,
allowed by rotational and time-reversal symmetry:
\be
\Hone =A\frac{p^4}{4\mu^3}+ B\frac{\alpha}{\mu^2}\frac{p^2}{r}+ C\frac{\alpha}{\mu^2}\frac{p_r^2}{r}+ D \frac{\alpha^2}{\mu r^2} +
2\pi \alpha E \delta^3(r), \label{H1PN}
\ee
where $p_r \equiv\frac{1}{2}\left(\frac{\vec{r}}{r}{\cdot}\vec{p}+\vec{p}{\cdot}\frac{\vec{r}}{r}\right)$ is the radial derivative.
To be fully precise, including operator ordering choice in the quantum theory (in which we will also be interested below),
the products here are defined as follows: $\frac{p^2}{r} \equiv \frac12(p^2\frac{1}{r}+\frac{1}{r}p^2)$
and similarly for $\frac{p_r^2}{r}$.

We stress that we do not include spin nor fermion exchanges. This would be
necessary to reproduce the full supermultiplet structure of the bound states (see \cite{Herzog:2009fw} for the $\mathcal{N}=1$ case),
but since we already know the supersymmetry degeneracies it suffices to study one convenient state within the multiplet.
As in ref.~\cite{Caron-Huot:2014gia},
we choose a state with two identical complex scalars as defined above eq.~\eqref{eq:amp phi}.  

It is also important to note that eq.~(\ref{H1PN}) is \emph{not} the result of integrating out
the gravitational field, but only those components which have energy above the typical binding energy $\sim p^2/(\mu)$.
The surviving gravity modes would be called ultrasoft in the QCD literature, and associated with so-called tail effects in gravity,
see \cite{Brambilla:2004jw,Porto:2016pyg}; however they are not important at the order to which we are working.

Let us briefly review how the various coefficients can be obtained by EFT matching with the scattering
amplitudes. The coefficient $A$ is obtained simply by matching
the relativistic kinetic energy of the two bodies (as measured when they are infinitely far apart):
\be
 \frac{p^2}{2\mu} + A\frac{p^4}{4\mu^3} +\ldots =
 \left(\frac{p^2}{2m_A}- \frac{p^4}{8m_A^3} +\ldots\right)
 +\left(\frac{p^2}{2m_B} -\frac{p^4}{8m_B^3}+\ldots\right) 
\ee
which gives the familiar result usually written in the gravity literature as \cite{Buonanno:1998gg}
\be
 A = -4\mu^3\left(\frac{1}{8m_A^3}+\frac{1}{8m_B^3}\right)=\frac{3\nu-1}{2}, \quad\mbox{with}
 \quad \nu\equiv \frac{m_Am_B}{(m_A+m_B)^2}. \label{def A}
\ee
The coefficients $\alpha$, $B$ and $C$ follow from matching with the scattering amplitude at first (leading) order in $G_N$,
which the EFT Hamiltonian predicts as
\be\label{first born}
\begin{aligned}
\tilde{\mathcal{A}}^{(1)}_{\phi\phib\phi\phib}\Big|_{EFT} &=
-\langle p'| \left(-\frac{\alpha}{r} +B\frac{\alpha}{\mu^2}\frac{p^2}{r}+ C\frac{\alpha}{\mu^2}\frac{p_r^2}{r} +2\pi \alpha E \delta^3(r)+ O(\mbox{2PN})\right) |p\rangle
\\
&=\frac{4\pi \alpha}{q^2}\left(1 - (B+C) \frac{p^2}{\mu^2}\right) -2\pi (C+E)
+ O(\mbox{2PN})
\end{aligned}
\ee  
where superscript $(1)$ simply means first order in the coupling constant, $\vec{q}=\vec{p'}-\vec{p}$ and the second line follows simply from computing Fourier transforms.
These are detailed in eq.~(\ref{H1 matrix element}), in which we have further set $p'^2=p^2$ using energy conservation.
Note that in this section we use non-relativistic normalization, $\langle p'|p\rangle=(2\pi)^3\delta^3(p-p')$ (whence the tilde on the left-hand-side),
while the sign comes from the usual Born approximation
$i\mathcal{A}\approx -iH$. The Fourier transforms were greatly simplified by using the following identity,
which holds for the mentioned choice of operator ordering:
\be
\frac{p_r^2}{r} = \frac{p^2}{r} + \frac14 [p^2,[p^2,r]] + 2\pi \delta^3(r). \label{pr identity}
\ee
Matching eq.~(\ref{first born})
with the velocity expansion of the tree result in eq.~(\ref{a00}) gives
\be\begin{aligned}
\label{alphaBC}
 \alpha &= G_Nm_Am_B \times 2(1-\cos\phi_1)(1-\cos\phi_2), \\
 B+C &= -\nu-\frac{1}{2}\frac{1-\cos\phi_1\cos\phi_2}{(1-\cos\phi_1)(1-\cos\phi_2)}.
\end{aligned}\ee
In particular, the Newtonian potential for our system is not $G_Nm_Am_B$ but contains an extra factor
proportional to the charge misalignment, as anticipated.
For instance, when the charges are maximally misaligned, the total force is eight times stronger than that from gravity alone.
We also note that only the combination $B+C$ is fixed by the on-shell amplitude.
This reflects a gauge redundancy associated with a coordinate transformation (of the form $r\to r(1+ c\frac{G_Nm}{r})$, see \cite{BjerrumBohr:2002kt,Hiida:1972xs}),
which allows to move information between $B$ and $C$.
The contact interaction $E+C$ is in principle fixed at tree-level, but cannot be reliably obtained from the cut diagrams
considered in the preceding section without further assumptions.

To find $D$, we need to compare the one-loop amplitude from the preceding subsection
with the $\alpha^2$ expansion of the amplitude computed in the EFT.
Perturbing around the theory with $\alpha=0$ (as appropriate for scattering states),
this receives contributions from both the first and second Born approximation:
\begin{eqnarray}
\tilde{\mathcal{A}}^{(2)}_{\phi\phib\phi\phib}\Big|_{EFT}
&=& 
-\langle p'|H^{(2)}|p\rangle+\int \frac{d^3\ell}{(2\pi)^3} \langle p' |H^{(1)}|\ell \rangle\frac{1}{E_{\ell}-E_p-i0}\langle\ell|H^{(1)}|p\rangle
\label{first and second born}
\end{eqnarray}
where the superscripts on $H$ denote the powers of $\alpha$,
and $E_p = \frac{p^2}{2\mu} + A\frac{p^4}{4\mu^3}$.
The second term requires performing a (three-dimensional) loop integral, as detailed in appendix \ref{app:fourier};
the total to 1PN accuracy is:
\be\begin{aligned}
\label{EFT A1}
\tilde{\mathcal{A}}^{(2)}_{\phi\phib\phi\phib}\Big|_{EFT} &=
-\frac{2\pi^2\alpha^2}{\mu\sqrt{-t}}\times 
\underbrace{\left[A+2B+C+D\right]}_{f_0(A,B,C,D)}+
\frac{4\pi\alpha^2\mu}{\sqrt{-p^2-i0}}\left(\frac{1-f_1 \frac{p^2}{\mu^2}}{q^2}\log \frac{q^2}{\bar\mu^2}
+\frac{C+E}{\mu^2}\log \frac{-p^2}{\bar\mu^2} \right)
\end{aligned}
\ee
with $f_1=A+2B+2C$. 
The first term contains the important new information as it gives the precise subtractions needed
to extract the yet unknown coefficient $D$ from
the $1/\sqrt{-t}$ term in the amplitude.
In particular, from the no-triangle property discussed at the end of subsection \ref{oneloopbhscattering}, we find that
\be
\label{noypole}
 f_0 = A+2B+C+D =0.
\ee
The second term in eq.~(\ref{EFT A1}) gives no new information since
it contains only the constants $A, B+C$ and $E+C$,
already determined at lower order.
It's insensitivity to $H^{(2)}$ is reflected in fact that it is non-analytic in the energy $E\sim p^2$.
This term serves as a valuable consistency check on the structure of the effective theory, and we have verified that it is precisely as predicted by \eqref{a2}.

\subsubsection{Post-Minkowski expansion and integrand-level subtractions}
\label{ssec: integrand subtractions}

As this manuscript was being completed, a very efficient method to perform the EFT matching appeared \cite{Cheung:2018wkq}
(see also \cite{Bjerrum-Bohr:2018xdl}).
It allows to perform integrand-level subtractions of the lower-order iterations, dramatically simplifying the integrals to be computed.
In fact one gets for free the full post-Minkowski Hamiltonian: one keeps the full dependence on velocity, to a given order in $G_N$
\cite{Damour:2016gwp}.
For future reference, in this section we record the corresponding expressions; the reader interested in the precession can skip to subsection \ref{ssec:precession}.

The authors of \cite{Cheung:2018wkq} provided a unique map between on-shell scattering amplitudes and classical potentials. They overcame gauge ambiguities by eliminating the terms in the Hamiltonian using the equation of motion, thus leaving a Hamiltonian which depends on the same variables as the scattering amplitudes.
This choice of gauge in our setup is equivalent to putting $C=0$, and therefore $B=B+C$.
We restrict our focus to first and second order in the coupling constant in most of this discussion.
But the argument woks in general, see \cite{Cheung:2018wkq}. In this gauge the full post Minkowskian Hamiltonian takes the form:
\begin{equation}
H=\Bigg(\frac{p^2}{2\mu}+A\frac{p^4}{4\mu^3}+\ldots\Bigg)+ V \label{H Cheung}
\end{equation}
where V is the classical potential. The gauge is fixed by requiring that matrix elements of V
depend only on the average of the final and initial momenta, and momentum transfer $q=|\vec{p}'-\vec{p}|$:
\begin{equation}
\langle p^{\prime}|V|p\rangle\equiv V\left(\frac{p^2+p'^2}{2}, q\right),\qquad
V(p^2,q) = \frac{c_1(p^2)}{q^2} + \frac{c_2(p^2)}{|q|}+\ldots \label{V gauge}
\end{equation}
The results just obtained give for example the velocity expansions:
\be\begin{aligned}
\label{cis}
c_1&=-4\pi\left(\alpha-\alpha(B+C)\frac{p^2}{2\mu^2}+O(p^4)\right),\\
c_2&= \frac{2\pi^2\alpha^2 D}{\mu^2}+O(p^2),
\end{aligned}\ee
where the coefficients $A$, $B+C$ and $D$ can be found in eqs.~(\ref{def A}), (\ref{alphaBC}), (\ref{noypole}).
On the other hand, in the gauge (\ref{V gauge}) we can read off the full velocity dependence of $c_1$ just by
matching with the tree-level amplitude given in eq.~\eqref{result Atree 2-angles}:
\begin{equation}
c_1(p^2)=-\frac{16G_N\pi m_A^2m_B^2\times 2(\cosh\eta_{AB}-\cos \phi_1)(\cosh\eta_{AB}-\cos \phi_2)}{4E_AE_B}
\end{equation}
where $\cosh\eta_{AB} = \frac{E_AE_B+p^2}{m_Am_B}$ and $E_i=\sqrt{p^2+m_i^2}$.
It is trivial to see that the velocity expansion of $c_1(p^2)$ matches with eq.~\eqref{cis}.

The more technically involved part of our previous EFT matching calculation was the calculation of the second-order
Born term in eq.~\eqref{second born}. For the effective Hamiltonian (\ref{H Cheung}), the corresponding term
can be written in the form
\begin{equation}
\tilde{\mathcal{A}}_{\text{EFT}}^{\rm 1-loop}=-\int_{\ell}\frac{\mathcal{N}_{\text{EFT}}^{1-\text{Loop}}}{X_1^2X_2^2Y_1}
\end{equation}
with $\int_\ell \equiv \int \frac{d^3\vec{\ell}}{(2\pi)^3}$ and 
\begin{equation}
X_1=|p-\ell|, \quad X_2=|\ell-p^{\prime}|, \quad Y_1=\ell^2-p^2.
\end{equation}
Physically, $Y_1$ measures the off-shellness of matter propagators.
Since the potential (\ref{V gauge}) depends on velocity through the combination $\frac{p^2+\ell^2}{2} = p^2 + \frac12Y_1$,
the numerator is a polynomial in $Y_1$. However, terms higher than linear can be neglected in the classical limit.
To linear order, the EFT integrand is thus given as \cite{Cheung:2018wkq}:
\begin{equation}
\label{NEFT}
\begin{split}
\mathcal{N}_{\text{EFT}}^{\rm 1-loop}=\left[\frac{2E_AE_B}{(E_A+E_B)}+Y_1\left(\frac{1-3\xi}{2(E_A+E_B)\xi}+(E_A+E_B)\xi\partial_{p^2}\right)\right]c_1^2(p^2)
\end{split}
\end{equation}
where $\xi=\frac{E_AE_B}{(E_A+E_B)^2}$ which is the post Minkowskian analogue of $\nu$. The post-Newtonian expansion of this is easily deducible from eq.~\eqref{EFTintegrand}:
\begin{equation}
\label{NEFTpn}
\mathcal{N}_{\text{EFT}}^{\rm 1-loop}=2\mu(4\pi\alpha)^2\left(1-(A+2B)\frac{p^2}{\mu^2}-\frac{A+2B}{2\mu^2}Y_1\right)+O(p^4)
\end{equation}
The next step of the matching procedure of  \cite{Cheung:2018wkq} is to write the integrand of the full theory in the same form.
For a generic theory, using the scalar integral decomposition, the exact 1-loop amplitude can be written as:
\begin{equation}
\begin{split}
\mathcal{A}^{\rm 1-loop}=d_{\Box}\mathcal{I}_{\Box}+d_{\lhd}\mathcal{I}_{\lhd}+d_{\rhd}\mathcal{I}_{\rhd}+\ldots
\end{split}
\end{equation}
Here the dots denotes terms that do not contribute classically. After integrating over the energy $\ell^0$,
the scalar basis integrals reduce to three dimensional integrals of the above form,
\begin{equation}
\label{boxtriangle}
\mathcal{I}_{\Box}=\frac{i}{2(E_A+E_B)}\int_\ell\frac{1}{X_1^2X_2^2Y_1}+\ldots,\qquad
\mathcal{I}_{\rhd}=-\frac{i}{4m_A}\int_\ell\frac{1}{X_1^2X_2^2}+\ldots
\end{equation}  
where we have retained the triangle integrals for pedagogical reason, even though they have zero coefficients in $\N$. Here the dots denote  quantum mechanical contributions. In eq.~\eqref{boxtriangle} we explicitly see that the box integral has a simple pole in $Y_1$
(this integral is IR divergent and was computed in eq.~\eqref{a2}).
Now we have all the ingredients for matching the calculation at the integrand level:
\begin{equation}
\begin{split}
0=\tilde{\mathcal{A}}^{\rm 1-loop}-\tilde{\mathcal{A}}^{(2)}_{\text{EFT}}=\left[\int_\ell \frac{\mathcal{N}_{\text{EFT}}^{\rm 1-loop}-\mathcal{N}^{1-\text{Loop}}}{X_1^2X_2^2Y_1}-\frac{c_2(p^2)}{q}\right].
\end{split}
\end{equation}
Using eq.~\eqref{boxtriangle}, one can see that the scalar box integral in the full theory simply cancels
the term without $Y_1$ in eq.~\eqref{NEFT}. Thus we only need to compute integrals that do not involve $Y_1$ in the denominator!
There is only one, recorded already in appendix \ref{app:fourier}:
\begin{equation}
\int_{\ell}\frac{\mathcal{N}(p^2)}{X_1^2X_2^2}=\frac{\mathcal{N}(p^2)}{8q}.
\end{equation}
This gives us the full velocity dependence of the first post-Minkowski potential in $\mathcal{N}=8$ supergravity:
\begin{equation}
\begin{split}
c_2(p^2)=\frac{1}{128E_AE_Bq}\left[
\left(\frac{8E_AE_B(1-3\xi)}{(E_A+E_B)\xi}+16E_AE_B(E_A+E_B)\xi\partial_{p^2}\right)c_1^2(p^2)+
\frac{d_{\rhd}}{m_A}+\frac{d_{\lhd}}{m_B}\right],
\end{split}
\end{equation}
where we recall from the preceding section that the triangle coefficients vanish: $d_{\lhd}=d_{\rhd}=0$.

Performing the velocity expansion, using eq.~(\ref{cis}), we can write $c_2(p^2)$ as
\begin{equation}
c_2(p^2)=\frac{-2\pi^2\alpha^2}{\mu}(A+2B) \quad \Rightarrow \quad A+2B+D=0,
\end{equation}
which precisely reproduces our earlier finding in eq.~(\ref{noypole}), now specialized to the gauge $C=0$.
In general, the right-hand-side would be proportional to the triangle coefficient.
The upshot is that since this method operates at integrand-level, it should enable efficient calculations at higher loops.

\subsection{Precession from the amplitude}\label{ssec:precession}

Given the 1PN Hamiltonian, eq.~\eqref{H1PN}, it is a straightforward exercise
in perturbation theory to work out the perihelion precession rate.
Following \cite{Nabet:2014kva}, we will adopt an approach which eschews explicit knowledge
of the Newtownian orbits but rather focuses on symmetries, namely the Laplace-Runge-Lenz (LRL) vector
\be
 \vec{A}_{\rm 0PN} = \vec{p}\times\vec{L} - \mu \alpha\frac{\vec{r}}{r}
\ee
where $\vec{L}=\vec{r}\times \vec{p}$ is the angular momentum.
It is easy to check that it commutes with respect to $H_{0PN}$ given in eq.~\eqref{H0PN}.
It is also easy to check that it does not commute with its first post-Newtonian correction.
However, this does not necessarily imply a precession, because some terms may average to zero
over one orbit. Instead of explicitly computing the average of $\dot{\vec{A}}=-i [\vec{A}_{\rm 0PN},H_{\rm 1PN}]$,
a somewhat simpler but equivalent procedure is to average $H_{\rm 1PN}$ itself.

This can be done conveniently by writing terms as much as possible
in terms of commutators with $\Hzero$, whose time-average trivially vanish, and powers of $\Hzero$,
which commute with $\vec{A}_{\rm 0PN}$.
For example, the identity in eq.~(\ref{pr identity}) admits a variant which is more suitable for bound orbits,
where we perturb around the full leading Hamiltonian $\Hzero$:
\be
 \frac{p_r^2}{r} = \frac{p^2}{r} + \mu^2[\Hzero,[\Hzero,r]] -\frac{\mu\alpha}{r^2} +2\pi\delta^3(r).
\ee
The second term averages to zero, while the rest are terms already present in the Hamiltonian.
To simplify the other terms, we use that the following, being a commutator, averages to zero (the so-called virial theorem):
\be
 i[\Hzero,\vec{p}{\cdot}\vec{r}] = \frac{p^2}{\mu} - \frac{\alpha}{r} = 2\Hzero + \frac{\alpha}{r}.
\ee
To simplify $\frac{p^2}{r}$ itself, we write $p^2$ using the Hamiltonian and then use the preceding formula:
\be
\frac{\alpha}{2\mu} \frac{p^2}{r} \equiv \left(\frac{\alpha}{r}+\Hzero \right)\frac{\alpha}{r}
=\frac{\alpha^2}{r^2}+ i [\Hzero, \Hzero \vec{p}{\cdot}\vec{r}]-2\Hzero^2.
\ee
In this way we can eliminate all nontrivial averages from the 1PN Hamiltonian ~\eqref{H1PN}
up to a single one, $1/r^2$:
\be \label{Hone nice}
 \Hone = -\frac{\Hzero^2}{\mu}(3A+4B+4C) +i[\Hzero,X]
 + \frac{\alpha^2}{\mu r^2}
 \underbrace{\left(A+2B+C+D\right)}_{f_0(A,B,C,D)} +2\pi\alpha(C+E)\delta^3(r)
\ee
where $X=\frac{1}{\mu}\{ H_0,p{\cdot}r\} (A+B+C) -iC [\Hzero,\alpha r]$.
This decomposition is advantageous because the first term commutes with the LRL vector, the second term averages to zero,
and the last term vanishes classically, so the $f_0$ term gives directly the time-averaged precession:
\be
 \langle \dot{\vec{A}}\rangle_{\rm cl} \equiv -i [\vec{A}_{\rm 0PN},\langle \Hone\rangle ]
 \simeq f_0(A,B,C,D)\times [\vec{A}_{\rm 0PN},-i \langle\frac{\alpha^2}{\mu r^2}\rangle]. \label{precession classical}
\ee
We see that the classical precession is controlled by the same combination $f_0(A,B,C,D)$
which controls the $1/\sqrt{-t}$ coefficient in the one-loop scattering amplitude (\ref{EFT A1}).
This immediately implies that there is no classical precession for half-BPS black holes in $\mathcal{N}=8$ supergravity,
and that this property is equivalent to the absence of scalar triangles noted in ~\eqref{box integrals}.

Physically, the first term in eq.~(\ref{Hone nice}) could be absorbed into a redefinition of the energy;
this appears to be at the heart of the so-called effective one-body parametrization. After this,
the only remaining classical physical effect is the $\alpha^2/(\mu r^2)$ term which can be absorbed into an effective metric;
see \cite{Buonanno:1998gg,Damour:2008qf} for the generalization of this idea to higher orders.

For our purposes, the commutator term gives a simple way to redefine the LRL vector
so that it is conserved. Namely, we add a post-Newtonian correction to it:
\be
 \vec{A}_{\rm 1PN} = i [\vec{A}_{\rm 0PN},X],
\ee
which gives exactly:
\be
\label{LRL corrected}
\begin{aligned}
 ~\langle[\vec{A},H]\rangle_{\rm full} &\equiv [\vec{A}_{\rm 0PN},\langle\Hone\rangle] + [\langle\vec{A}_{\rm 1PN}\rangle,\Hzero] + O(2{\rm PN})
\\ &= f_0(A,B,C,D) [\vec{A}_{\rm 0PN},\langle\frac{\alpha^2}{\mu r^2}\rangle] + 2\pi \alpha(C+E)[\vec{A}_{\rm 0PN},\langle \delta^3(r)\rangle] .
\end{aligned}\ee
The first coefficient on the last line vanishes in the case of $\mathcal{N}=8$ supergravity, while the second coefficient affects only the $s$-wave states
(zero angular momentum).
Classically, this shows that a conserved LRL vector continues to exist in $\N$ including the leading relativistic corrections.
(For an elegant relativistic form of the LRL vector in a different setup, see \cite{Alvarez-Jimenez:2018lff}.)
At the quantum level, this shows that
the SO(4) symmetry responsible for the Hydrogen-like quantum mechanical degeneracies in the energy levels is
preserved (at least for angular momenta $L>0$),
ensuring that the bound states in $\mathcal{N}=8$ supergravity retain their degeneracies at 1PN order ($H\sim \mu\alpha^4$).
The degeneracy extends to $L=0$ if and only if the contact interaction coefficient $C+E=0$;
the method of the preceding section does not allow to reliably compute this coefficient.

As a validation of these results, we can work out quantitatively the precession in other theories where $f_0\neq 0$.
The average in eq.~(\ref{precession classical}) can be written as (see for example \cite{Nabet:2014kva}, section 5),
$\langle\frac{\alpha^2}{\mu r^2}\rangle = \alpha (-2\mu H_{0PN})^{3/2}/[\mu^2L]$, where L is the magnitude of the conserved angular momentum vector. 
Using Leibniz rule we have, classically, $[\vec{A},-i/L]=(\hat{L}\times \vec{A})/L^2$, so the precession rate (\ref{precession classical})
can be written as
\be
\langle \dot{\vec{A}} \rangle _{\rm cl}= f_0(A,B,C,D)\frac{\alpha}{\mu^{2}L^2}
\left(-2\mu H_{0PN}\right)^{\frac{3}{2}} (\hat{L}\times\vec{A}).
\ee
Multiplying by the period $T=2\pi \mu^2\alpha/(-2\mu H_{0PN})^{3/2}$ gives the following angular precession per orbit:
\be
 \Delta\phi = -\frac{2\pi\alpha^2}{L^2}f_0.
\ee
Comparing with eq.~(\ref{EFT A1}) we can express this directly
in terms of a coefficient in the one-loop amplitude:
\be
\label{precessionfromEFT}
\Delta\phi = \frac{\mu}{\pi L^2} \tilde{\mathcal{A}}^{\rm 1-loop}\Big|_{\frac{1}{\sqrt{-t}}}.
\ee
This is the main result of this section:
it expresses the 1PN precession, in any theory. in terms of the 1-loop on-shell scattering amplitude;
it vanishes in $\N$ because triangles are absent.
A more direct derivation (by-passing an effective Hamiltonian) would be interesting.
As a simple validation, we can check that this reproduces the celebrated result in General Relativity, taking
the coefficient of $1/\sqrt{-t}$ in the amplitude in eq.~(\ref{gramp}):
\be
 (\Delta\phi)_{\rm GR} = \frac{6\pi (G_Nm_Am_B)^2}{L^2}.
\ee

\section{Probe limit}
\label{probe}

An important question now is to determine whether the lack of precession just calculated to first post-Newtonian order survives to higher orders and beyond perturbation theory. For getting a hold of the answer, we turn to the probe limit, $m_B \to 0$, where we can study bound orbits to all orders in velocity, in terms of geodesic motion in a classical background solution.
This will be most conveniently done by uplifting the $\N$ theory to type IIA supergravity compactified on a T${}^6$.

For the heavy background $m_A$, we choose a stack of D6 branes, which preserves all the symmetries of the torus.
Using a D6 probe with various U(1) flux, we will be able to realize the general case of three angle
charge misalignment described in section \ref{setup}.
As a warm-up, we first consider D2 and D0 branes, the latter being magnetically charged with respect to the D6.
We will find no precession to all orders in velocity, in all cases.
The central charges of these various probes are worked out in appendix \ref{app: charges}.

\subsection{D2 probe in D6 background}

For pedagogical reasons, let us start with a D2 brane probe, since it interacts neither magnetically nor electrically with
the D6 background. This will realize a single free misalignment angle, corresponding to the probe's Kaluza-Klein momentum.
As we will see shortly, there is no Newtonian force in this case, and no closed orbits, 
but it is nonetheless an instructive starting point.
The DBI action in this case is:
\begin{equation}
\label{D210Daction}
        S_{D2}=T_{D2}\int d^3\xi e^{-\Phi}\sqrt{-\det(\gamma_{ab})}
\end{equation}
where $\gamma_{ab}$ is the induced metric on the brane.
The background metric induced by the D6 stack, see~\cite{Aharony:1999ti}, is a BPS solution carrying electric charge with respect to $A_7$:
\begin{equation}
\label{g}
      ds^2=-\frac{1}{\sqrt{H(r)}}dt^2+\sqrt{H(r)}(dr^2+r^2d\Omega^2)+\frac{1}{\sqrt{H(r)}}\sum_{i=1}^6dx^idx^i
\end{equation}
where 
\begin{equation}
    H(r)=1+\frac{4G_Nm_A}{r}\,, \qquad G_N=\frac{g^2_sl_s^2}{8}\left(\frac{(2\pi l_s)^6}{{\rm Vol}({\rm T}^6)}\right).
\end{equation}
Here the dilaton field (we absorbed the string coupling into $T_2$ so that $\Phi$ vanishes at infinity) is given as
\begin{equation}
    e^{\Phi}=H(r)^{\frac{-3}{4}}.
\end{equation}
Since the background metric is translation invariant along the T${}^6$, we can look for solutions where the D2
spans only two directions of the torus and is supported on a point in the four other directions.
Two integrations can then
be done trivially, so the action effectively becomes the action of a point particle in a gravitational background
 \begin{equation}
        S_{D2}=m_B\int d\lambda\sqrt{\tilde{g}_{mn}\partial_{\lambda}X^m\partial_{\lambda} X^n}
\end{equation}
where the effective metric, $\tilde{g}_{mn}=e^{-2\Phi}g_{\bot\bot}g_{\bot\bot}g_{mn}$ is explicitly:
\begin{equation}
d\tilde{s}^2=-dt^2+H(r)(dr^2+r^2d\Omega^2)+\sum_{i=1}^6dx^idx^i\,. \label{metric D2}
\end{equation}
Let us now briefly review the standard procedure from this point.
Now one can choose a specific parametrization $\lambda=\tau$, along which:
\begin{equation}
\label{gm}
\tilde{g}_{mn}\frac{dx^m}{d\tau}\frac{dx^n}{d\tau}=-1.
\end{equation}
Taking the conserved angular momentum to be in the $z$ direction,
the quantities conserved by the equations of motion are then:
\begin{equation}
\label{cons0}
    \begin{split}
        &E=-\tilde{g}_{00}\frac{dt}{d\tau},\\
        &P_{\bot i}=\tilde{g}_{ii}\frac{dx^i}{d\tau} \qquad (i=4\ldots,9),\\
        &P_{\phi}=\tilde{g}_{\phi\phi}\frac{d\phi}{d\tau}=\tilde{g}_{rr}r^2\sin^2\theta_0 \frac{d\phi}{d\tau}.
        \end{split}
\end{equation}
Here $\theta_0$ is a constant, equal to $\theta_0=\pi/2$ for the moment since the motion is in the equatorial plane,
but we introduce it for future use below.
Now we can rewrite eq.~\eqref{gm} as a function of conserved variables eq.~\eqref{cons0}:
\begin{equation}
\frac{E^2}{\tilde{g}_{00}}+\tilde{g}_{rr}\left(\frac{dr}{d\tau}\right)^2+\frac{P_{\phi}^2}{\tilde{g_{\phi\phi}}}+\frac{P_{\bot i}^2}{\tilde{g}_{\bot\bot}}=-1.
\end{equation}
Using this formula to solve for $\frac{d\phi}{dr}$ and integrating it gives the angular change over a given change of $r$:
\begin{equation}
\label{prec0}
\Delta\phi=\int^{r_{\max}}_{r_{\min}} dr\frac{d\phi}{dr}
=\int^{r_{max}}_{r_{min}}\frac{dr/r}{\sqrt{\frac{((E^2-P_{\bot}^2)\tilde{g}_{rr}(r)}{\tilde{g}_{00}(r)P_{\phi}^2}r^2\sin^4\theta_0-\frac{\tilde{g}_{rr}r^2\sin^4\theta_0}{P_{\phi}^2}-\sin^2\theta_0}}.
\end{equation}
Setting $\sin\theta_0=1$, this reduces to:
\begin{equation} \label{no precession D2}
\begin{split}
\Delta\phi=&\int_{r_{\min}}^{r_{\max}}\frac{dr/r}{\sqrt{\frac{(E^2-P_{\bot}^2)}{P_{\phi}^2}H(r)r^2-\frac{H(r)}{P_{\phi}^2}r^2-1}}
\end{split}
\end{equation}
Notice that the argument of the square root is a quadratic polynomial in $r$ (because $H(r)=1+\frac{4G_Nm_A}{r}$ is linear in $1/r$),
and so it has two roots.  For a closed orbit, these two roots $r_{\min}$ and $r_{\max}$ would be positive, and the integral could be rewritten
as a contour integral, as will be done in the other examples below.
In the case of eq.~\eqref{no precession D2} it is easy to see however that the roots will never be both positive -
there are no closed orbits.


\subsection{D0 probe in D6 background}

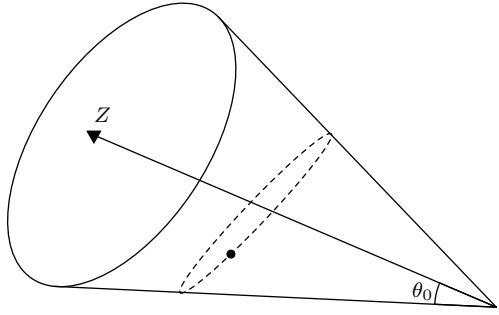
\begin{figure}[t]
    \centering
    \begin{tikzpicture}[line cap=round,line join=round,>=triangle 45,x=1cm,y=1cm]

\clip(-1.4555445059655316,-5.47949466243121) rectangle (8.64807693695515,0.26622053937752);
\draw [shift={(7.4,-4.1)},line width=0.5pt] (0,0) -- (156.82726498076693:0.8192125494260012) arc (156.82726498076693:177.30304183133168:0.8192125494260012) -- cycle;
\draw [rotate around={-123.86762695755795:(2.420349019594798,-1.941434874663692)},line width=0.5pt] (2.420349019594798,-1.941434874663692) ellipse (2.141545929824247cm and 1.1274108975249064cm);
\draw [line width=0.5pt] (1.523885878012233,-3.8232021481778427)-- (7.4,-4.1);
\draw [line width=0.5pt] (7.4,-4.1)-- (3.7032682639764154,-0.23683606846806105);\draw [line width=0.5pt] (7.4,-4.1)-- (2.05,-1.81);\draw [rotate around={-133.54613098517555:(4.2016463324144055,-2.850893967086937)},line width=0.5pt,dash pattern=on 2pt off 2pt] (4.2016463324144055,-2.850893967086937) ellipse (1.4578079728369195cm and 0.1491503024529874cm);
\begin{scriptsize}
\draw [fill=black,shift={(2.05,-1.81)},rotate=180] (0,0) ++(0 pt,3pt) -- ++(2.598076211353316pt,-4.5pt)--++(-5.196152422706632pt,0 pt) -- ++(2.598076211353316pt,4.5pt);
\draw[color=black] (2.162644253999307,-1.5399597042442377) node {$Z$};\draw [fill=black] (3.8718302635826105,-3.393775619534346) circle (1.5pt);
\draw[color=black] (6.4219668183327465,-3.889140672675301) node {$\theta_0$};\end{scriptsize}\
\end{tikzpicture}
    \caption{The orbit of an electrically and magnetically charged object is confined to the surface of a cone with half-angle $\theta_0$.}
    \label{fig:cone}
\end{figure}

As a second step toward the general case,
let us now consider the case of a D0 brane, which carries magnetic charge with respect to the $A_7$ RR-form
sourced by the D6 (it is dual to the $A_0$ field to which the D0 couples, $dA_{7}=\ast dA_0$).

The DBI action for the D0 brane can be written as:
 \begin{equation}
        S_p=m_B\int d\lambda e^{-\Phi}\sqrt{-g_{mn}\partial_{\lambda}X^m\partial_{\lambda}X^n}+N_0\int A_m\frac{dx^m}{d\lambda}d\lambda
\label{action D0}
\end{equation}
where $m_B$ is the mass and $N_0$ is the number of D0 brane probes. 
The vector potential takes the familiar Dirac monopole form
\begin{equation}
\label{A}
    \vec{A}(r)=\frac{N_6}{2r}\frac{[\vec{r}\times \vec{n}]}{r-(\vec{r}\cdot\vec{n})}
\end{equation}
where $N_6$ is the number of D6 branes.
Note that the vector potential is nonvanishing only along the four space-time directions $\vec{r}=(0,x,y,z)$, because the D6 background fills the T${}^6$.  
This has constant magnetic flux,
\begin{equation}
    \begin{split}
        \vec{B}=\vec{\nabla} \times \vec{A}=\frac{N_6}{2}\frac{\vec{r}}{r^3}
    \end{split}
\end{equation}
and the extremality constraint on the D0 and D6 brane masses and charges can be written as
\begin{equation}
 \label{bps}
 8G_Nm_Am_B= N_0N_6.
\end{equation}
This satisfies the Dirac quantization condition (\ref{NABdef}) with $\phi_i=\tfrac\pi2$.
Now in order to check whether we have precession or not, again, we find it useful to redefine the metric so $\tilde{g}_{mn}=e^{-2\Phi}g_{mn}$:
\begin{equation}
d\tilde{s}^2=-H(r)dt^2+H^2(r)(dr^2+r^2d\Omega^2)+H(r)\sum_{i=1}^6dx^idx^i\,.  \label{metric d2}
\end{equation}
Note that this differs from eq.~(\ref{metric D2}) by a factor $H(r)$: this was cancelled there by the volume of the two torus.
Following the same procedure as the D2 case, we can choose a parametrization such that eq.~\eqref{gm} holds for our metric.
The conserved quantities are then identical to eq.~(\ref{cons0}), with the exception of the angular momentum vector which
acquires an extra term:
\begin{equation}
\label{l}
    \begin{split}
    \vec{L}=\vec{\tilde{L}}+\frac{N_0N_6}{2}\frac{\vec{r}}{r}
    \end{split}
\end{equation}
where $\tilde{L}$ denotes the conventional angular momentum vector.
This is a well-known fact about magnetic charges (see for example \cite{Avery:2007xf,Balian:2005joa}) which can be easily verified using the equation of motion from the action (\ref{action D0}).
If we choose L to be along the z-direction we have:
\begin{equation}
\label{l1}
      \vec{L}\cdot\vec{r}=\frac{N_0N_6r}{2}\quad\Rightarrow\quad
        \frac{z}{r}=\frac{N_0N_6}{2L_z}\equiv \cos(\theta_0).
\end{equation}
This shows that the motion of the probe particle is confined to the surface of a cone with half-angle $\theta_0$ whose tip lies at the D6 stack,
see fig.~\ref{fig:cone}.
Substituting this relation back into eq.~(\ref{l}) then gives:
\begin{equation}
\label{lz}
   L_z=m_B\tilde{g}_{\phi\phi}\frac{d\phi}{d\tau}+\frac{(N_0N_6)^2}{4L_z} \quad\Rightarrow\quad
   P_\phi \equiv \tilde{g}_{\phi\phi} \frac{d\phi}{d\tau} =\frac{L_z\sin^2\theta_0}{m_B}.
\end{equation}
We can now calculate the precession angle over an orbit
for the D0-D6 system using the general formula \eqref{prec0} and the above expression
for $P_{\phi}$ in terms of the conserved quantity $L_z$:
\begin{equation}
\label{prec1}
    \begin{split}
         \Delta \phi=2\int_{r_{min}}^{r_{max}}dr\frac{d\phi}{dr}=&2\int_{r_{min}}^{r_{max}}\frac{dr/r}
      {\sqrt{\frac{m_B^2(E^2-P_{\bot}^2)}{L_z^2}H(r)r^2-\sin^2\theta_0-\frac{m_B^2H^2(r)r^2}{L_z^2}}}
    \end{split}
\end{equation}
Recall that $H(r)=1+\frac{4G_Nm_A}{r}$. Now:
\begin{equation}
\label{extrem}
\begin{split}
  &\frac{m_B^2(E^2-P_{\bot}^2-1)}{L_z^2}r^2+\frac{4G_Nm_Am_B^2(E^2-P_{\bot}^2-2)}{L_z^2}r-\frac{\overbrace{16G_N^2m_A^2m_B^2}^{\frac{(N_0N_6)^2}{4}}}{L_z^2}-\sin^2\theta_0=\\
 & \frac{m_B^2(E^2-p_{\bot}^2-1)}{L_z^2}r^2+\frac{4G_Nm_Am_B^2(E^2-P_{\bot}^2-2)}{L_z^2}r\underbrace{-\cos^2\theta_0-\sin^2\theta_0}_{-1}.
\end{split}
\end{equation}
On the first line we used the BPS condition \eqref{bps}.
Remarkably, the constant term in the polynomial is equal to $-1$, which implies that the orbits do not precess:
\begin{equation} \label{no precessionD0}
\begin{split}
\Delta\phi=&2\int_{r_{\min}}^{r_{\max}}\frac{dr/r}{\sqrt{\frac{m_B^2(E^2-p_{\bot}^2-1)}{L_z^2}r^2+\frac{4G_Nm_Am_B^2(E^2-P_{\bot}^2-2)}{L_z^2}r-1}}\\
=&\oint\frac{dr/r}{\sqrt{\frac{m_B^2(E^2-p_{\bot}^2-1)}{L_z^2}r^2+\frac{4G_Nm_Am_B^2(E^2-P_{\bot}^2-2)}{L_z^2}r-1}}=2\pi.
\end{split}
\end{equation}
The second line follows from the fact that the denominator is again quadratic in $r$:
this allows the integral over an orbit to be deformed into a contour integral around the pole at $r=0$.
We note the crucial role of the BPS conditions which relates the product of masses to the product of charges and therefore to the 
opening angle of the cone (specifically $\sin \theta_0$ term in eq.~\eqref{extrem}).
The fact that a magnetic monopole possess a conserved Laplace-Runge-Lenz vector 
is well-known in the non-relativistic limit (see \cite{Avery:2007xf,Anninos:2012gk}); here we found that the metric (\ref{metric d2}) leads
to no precession for a magnetic charge even for relativistic velocities.

\subsection{D6 probe with flux: general 3-angle case}

\label{d6probe}
Having gone through these two examples, we are ready to consider a configuration which realizes
the general case of three misalignment angles between the central charges.
This is realized by turning on U(1) flux on the D6 probe: this flux generates D0, D2 and D4 charges.
The DBI action in this case can be written as:
\begin{equation}
S_{DBI}=T_{D6}\int d^{7}\xi e^{-\Phi}\sqrt{-\det(\gamma_{ab}+2\pi\alpha^{\prime}F_{ab})}+\int_{p{\rm -brane}}C\wedge e^F
\end{equation}
Maintaining translation symmetry along the T${}^6$, we can turn on U(1) fluxes along 3 different planes:
$\int F_{45}dx_4dx_5=2\pi f_1$, $\int F_{67}dx_6dx_7=2\pi f_2$, $\int F_{89}dx_8dx_9=2\pi f_3$.
Integrating the DBI action along the 6 extra dimensions yields an action for a point particle in four dimensions,
\begin{equation}
S=m_B^{\prime}\int d\lambda e^{-\Phi}\sqrt{\prod_{i=1}^3(g_{\bot\bot}^2+f_i^2)g_{mn}\partial_{\lambda}x^m\partial_{\lambda}x^n}+f_1f_2f_3(N_6^{\prime})\int A_{m}\frac{dx^m}{d\tau}d\tau+q\int A_0 \frac{dt}{d\tau}d\tau
\end{equation}
Here $A_m$ is ($N_6$ times the Dirac monopole) vector potential given in eq.~\eqref{A},
and $A_0$ gives a static electric field from integrating the $C_7$ flux.
We can extract the (asymptotic) mass of the probe black hole from this equation to be:
\begin{equation}
\label{mb}
m_B=((1+f_1^2)(1+f_2^2)(1+f_3^2))^{\frac{1}{2}}m_B', \qquad m_B' = T_{D6}{\rm Vol}({\rm T}^6).
\end{equation}
Again we get an effective four-dimensional metric:
\begin{equation}
\tilde{g}_{mn}=e^{-2\Phi}g_{mn}\prod_{i=1}^3(g_{\bot\bot}^2+f_i^2).
\end{equation}
Note that the extremality constraint for two D6 branes can be written:
\begin{equation}
8G_N m_A m_B^{\prime}=N_6N_6^{\prime}\frac{{\rm Vol}({\rm T}^6)}{(2\pi l_s)^6}
\end{equation}
Proceeding as in the preceding two examples, the conserved quantities are:
\begin{equation}
\label{cons1}
    \begin{split}
        &E=-\tilde{g}_{00}\frac{dt}{d\tau}+\frac{q}{m_B'}A_0\\
        &P_{\bot i}=\tilde{g}_{ii}\frac{dx^i}{d\tau}=\frac{e^{-2\Phi(r)}}{\sqrt{H(r)}}\frac{dx^i}{d\tau} \qquad i=10-p,...,9\\
        &P_{\phi}=\tilde{g}_{\phi\phi}\frac{d\phi}{d\tau}=e^{-2\Phi}r^2\sin^2\theta_0\sqrt{H(r)}\frac{d\phi}{d\tau}
        \end{split}
\end{equation}
Since the electric potential makes the force vanish for static charges in the case of vanishing fluxes,
the Lagrangian in that limit must be $r$-independent we have simply
\begin{equation}
\frac{d}{dr}(\sqrt{\tilde{g}_{00}}+A_0)=0 \quad \Rightarrow\quad A_0=-\sqrt{\tilde{g}_{00}}\Big|_{f_1=f_2=f_3=0}.
\end{equation}
In addition, notice that we have a system of dyons,
so similar to the D0 case the motion is confined on the surface of a cone.
The cone angle and angular velocity are given as in eqs.~(\ref{l1}) and (\ref{lz}):
\be
  \cos\theta_0=\frac{N_6N^{\prime}_6f_1f_2f_3}{2L_z},\qquad
  P_\phi \equiv \tilde{g}_{\phi\phi} \frac{d\phi}{d\tau} =\frac{L_z\sin^2\theta_0}{m_B^{\prime}}.
\ee
Same as before, using eq.~\eqref{prec0} we can write:
\begin{equation}
\Delta\phi=2\int^{r_{max}}_{r_{min}} dr\frac{d\phi}{dr}=2\int^{r_{max}}_{r_{min}}\frac{dr/r}{\sqrt{\frac{((E-A_0)^2-P_{\bot}^2)\tilde{g}_{rr}(r)m_B^{\prime 2}}{\tilde{g}_{00}(r)L_z^2}r^2-\frac{\tilde{g}_{rr}r^2m_{B}^{\prime 2}}{L_z^2}-\sin^2\theta_0}}
\end{equation}
Remarkably, recalling that $H(r)=1+\frac{4G_N m_A}{r}$, we again have a quadratic polynomial in $r$ inside the square root!
Deforming again to get a contour integral around the $r=0$ pole, we have:
\begin{equation}
\begin{split}
\label{d6branenoprecession}
\Delta \phi=&\frac{2\pi i}{\sqrt{\left(\frac{(2\pi l_s)^6}{{\rm Vol}({\rm T}^6)}\frac{16f_1^2f_2^2f_3^2G_N^2m_A^2m_{B}^{\prime 2}}{L_z^2}-\sin^2\theta_0\right)}}\\
=& \frac{2\pi i}{\sqrt{-\cos^2\theta_0-\sin^2\theta_0}}=2\pi.
\end{split}
\end{equation}
Thus, thanks to the extremality condition, we find again that there is no precession and the orbits close, to all orders in velocity!

\section{Conclusion} \label{sec:conclusion}

In this paper, we studied the precession of orbits for bound states of two extremal (half-BPS) black holes in $\N$ supergravity.
In order to have interesting dynamics, we considered black holes that have misaligned electric and/or magnetic charge vectors, so that the static force between them does not cancel out. The general configuration of pairs of charge is labelled by three angles (see eq.~\eqref{second charge}). The precession of orbits is viewed as a diagnosis of whether this theory possesses hidden symmetries.
As a first step toward answering this question, we worked out the relevant on-shell scattering amplitude to one-loop accuracy
and use them to determine the precession to first post-Newtonian order.


Our main results are as follows. In eq.~(\ref{box integrals}) we observed that the one-loop amplitude for black hole scattering satisfies the no-triangle property, which was previously known for massless graviton scattering (see \cite{BjerrumBohr:2006yw}).
In section \ref{precession}, using standard effective theory methods to link the scattering problem to closed orbits,
we found that the no-triangle property is equivalent to absence of precession in the orbits (see eq.~\eqref{precessionfromEFT}).
A quantum version of the Laplace-Runge-Lenz (LRL) vector remains conserved, implying that the quantized energy levels
retain the degeneracies of the hydrogen atom, at least for states with nonzero angular momenta (see eq.~\eqref{LRL corrected}).
In section \ref{probe}, we extended the classical computation to all orders in the velocity in the probe limit $m_B\to 0$,
by uplifting the black holes to branes in string theory where we could use the Born-Infeld action.
Remarkably, even in the most general case involving electric and magnetic charges and arbitrary velocity, we found no precession (see eqs.~\eqref{no precessionD0} and \eqref{d6branenoprecession}).

We are thus motivated to make the following ``no-precession'' conjecture: that the energy levels of a pair of
black holes in $\N$ supergravity retain the degeneracies of the hydrogen atom to all orders in the perturbation theory.
These energy levels can be measured, for example, from the position of poles in the $2\to 2$ scattering amplitude of black holes.

Notice that the conjecture is formulated in terms of degeneracies rather than actual precession angle or conserved quantities.
This makes it meaningful at the quantum level. But this is also important classically:
because of radiation effects, which appear at higher orders in the post-Newtonian expansion and make the black hole pair an open system,
even the energy of the pair is not meaningfully conserved.
Similarly, there would be no reason to expect the LRL charge of the black holes to be conserved
unless one accounts for a possible LRL charge carried away by the radiation. While one might try to construct such charge,
looking at the spectrum avoids this difficulty.
Physically, scattering amplitudes avoid the complications of open systems because they answer exclusive questions, such as
the probability that the black holes scatter and \emph{not} radiate (or, emit a specific pattern of radiation).
The orbits not being stable means that the energy levels have an imaginary part (as defined from the position of S-matrix poles),
but the question of their degeneracies remains well-posed to all orders in $G_N$.%
\footnote{
At the nonperturbative level, the $S$-matrix (or even infrared-finite combinations of $S$-matrix elements
involving for example coherent states \cite{Bloch:1937pw,Strominger:2017zoo})
may not possess infinitely sharp poles, due to infrared radiation and resulting Sudakov logarithms,
which effectively spread poles over a nonperturbatively small window. 
Whether the degeneracies are well defined nonperturbatively, or only to all orders in the coupling, is a question we leave to future work.}

An important next question is to determine whether the conjecture holds true at the next post-Newtonian order, and what are its implications.
At one-loop, assuming the no-precession principle provides a symmetry explanation of the no-triangle property.
An interesting possibility is that the conjecture implies novel constraints on the integrand at two-loops and higher.
This question should be particularly tractable in the one-angle case, where the integrand is already known to high order
(being equal to that of ten-dimensional graviton scattering with nonvanishing Kaluza-Klein momenta, see discussion below eq.~(\ref{two-angles})); the integrand-level matching technique from ref.~\cite{Cheung:2018wkq} should also greatly simplify the calculation of precession,
as reviewed in section \ref{ssec: integrand subtractions}.
Other probes, such as the bending of light, could also be interesting to study.

Another key question is whether the LRL vector uplifts to give Ward identities for higher-point amplitudes --
in a similar way that the dual conformal symmetry of planar $\mathcal{N}=4$ super-Yang-Mills can be defined for arbitrary number of legs \cite{Drummond:2008vq} -- and whether the realization of $\N$ as double-copy of $\mathcal{N}=4$ helps construct this symmetry.
Finally, given that the calculations in section \ref{probe} were done at the level of ten-dimensional probes, one wonders whether some
more fundamental underlying symmetry exists already in ten dimensions.

\acknowledgments
Conversations with Alex Maloney, Emil Bjerrum-Bohr, Keshav Dasgupta and Maxim Emelin are gratefully acknowledged.
This work was supported in part by the National Science and Engineering Council of Canada,
the Canada Research Chair program, and the Fonds de Recherche du Qu\'ebec - Nature et Technologies.


\appendix

\section{Dimension of coset space}
\label{app:coset}

In this appendix, we show the details of how one can derive the parametrization of the double coset space introduced in \eqref{second charge}. First, Let's identity $SU(8)/H$ coset space. By acting with SU(8) generators on our canonical form,
\begin{equation}
C=
\begin{pmatrix}
0 & 1_{4\times 4} \\
-1_{4\times 4} & 0
\end{pmatrix}
\end{equation}

 we get:
\begin{equation}
    \tilde{C} =\sum_{i=1}^{63}(1+iT_i^T)C(1+iT_i)= C + i\sum_{i=1}^{63} \left(T_i^T\cdot C + i C\cdot T_i\right),
\end{equation}
where $T_i$s are the generators of the SU(8) algebra. Then by definition of the subgroup H, the points $\Tilde{C}$s belongs to the tangent space of the
$SU(8)/H$ and any element in the tangent space can be written in this way by the appropriate choice of parameters. These elements have the following form:

\begin{equation}
\label{indepcoeff}
  \tilde{C}=\centering
\begin{pmatrix}
A&B\\
B^{\star}&-A^{\star}
\end{pmatrix},
\end{equation}    

where A is complex and antisymmetric and B is traceless and antihermitian. Tracelessness of B comes from the tracelessness of the generators of SU(8). Now we need to figure out by acting with an H generator on this element which directions we can probe, in other words, what are our invariant angles:
\begin{equation}
    \Tilde{C}^{\prime} =\sum_{i=1}^{36}(1+iT_{H,i}^T)\Tilde{C}(1+iT_{H,i})= \Tilde{C} + i\sum_{i=1}^{36} \left(T_{H,i}^T\cdot \Tilde{C} + i \Tilde{C}\cdot T_{H,i}\right),
\end{equation}
where $T_{H,i}$ denote generators of H. Once we have the form of $\Tilde{C}^{\prime}$ it is easy to see how many directions in the coset space generators of H span. One just needs to build the matrix of the coefficients of each of the 36 generators of H, which is $36\times27$ dimensional (number of independent coefficient can be calculated from eq.~\eqref{indepcoeff} to be 27). The rank of this matrix is 24, thus there are 24 different direction that can be probed by H (these are the invariant angles) and 3 directions not accessible by H are the one parametrizing the angles of the double coset space H$\bs SU(8)/H$. \\
One can easily show that a generic element of the double coset space can be brought to the following form by the appropriate $SU(8)$ transformation.
\begin{equation}
C_B=m_B\times
\begin{psmallmatrix}
 &  &  &  & e^{i\phi_1} &  &  &  \\
 &  &  &  &  & e^{i\phi_2} &  &  \\
 &  &  &  &  &  & e^{i\phi_3} &  \\
 &  &  &  &  &  &  & e^{i\phi_4} \\
-e^{i\phi_1} &  &  &  &  &  &  &  \\
 & -e^{i\phi_2} &  &  &  &  &  &  \\
 &  & -e^{i\phi_3} &  &  &  &  &  \\
 &  &  & -e^{i\phi_4} &  &  &  & \\
\end{psmallmatrix}
,\qquad \prod_{i=1}^4 e^{i\phi_i}=1.
\end{equation}

\section{Black holes charges from brane configurations} \label{app: charges}

Here we compute the central charge of the black branes used in section \ref{probe} and extract the corresponding four-dimensional central charges. 
We can then study the misalignment angles between the charges in each setup.
The supersymmetry algebra for the extended objects is mentioned in \cite{deAzcarraga:1989mza} (see also \cite{Polchinski:1998rr})
and gives the anticommutator. 
\begin{equation}
    \{Q_{\sigma},Q_{\rho}\}=2(C\Gamma_{10}\Gamma^{\mu})_{\sigma\rho}P_{\mu}+2T(C\Gamma_{\mu_1\ldots\mu_p})T^{\mu_1\ldots\mu_p},
\end{equation}
where $\Gamma_{\mu_1\ldots\mu_p}$ is the antisymmetrized product of Dirac matrices in ten dimension and T is the p-volume tension of the extended objects. For the D6 brane stretched in directions $4\ldots 9$, for example,
\begin{equation}
     \{Q_{\rho},Q_{\sigma}\}=2(C\Gamma_{10}\Gamma_m)_{\rho\sigma}P^{m}+T_{D6}(C\Gamma_{4\ldots 9})_{\rho\sigma}.
\end{equation}
Dimensionally reducing to four dimensions, we decompose the ten-dimensional spinors and gamma matrices into tensor products
of four- and six-dimensional ones, for example
\begin{equation}
    \begin{split}
       & (C\Gamma_{10}\Gamma_m)_{\rho\sigma}=(c\gamma_{\mu})_{\alpha\beta}(C\Gamma_7\Gamma_7)_{AB}+c_{\alpha\beta}(C\Gamma_m\Gamma_7)_{AB},\\
       &(C\Gamma_{10})_{\rho\sigma}=(c)_{\alpha\beta}(C\Gamma_7)_{AB},\\
       &(C\Gamma_{4\ldots 9})_{\rho\sigma}=(c\gamma_5)_{\alpha\beta}(C\Gamma_7)_{AB}
    \end{split},
\end{equation}
where $\alpha,\beta=1,\ldots,4$ are SO(3,1) gamma matrices and $A,B=1,\ldots,8$ are SO(6) gamma matrix indices which are the $8\times 8$ skew symmetric central charges coming from KK momentum or charges of 10d extended object.
It is beneficial to enumerate these central charges to confirm that we can produce all 28 antisymmetric matrices.
Since Dp and D(6-p) branes are charged under the potentials dual to each other, $dA_{7-p}=\ast dA_{p+1}$, we have $C_{D0}=iC_{D6}$ and $C_{D2}=iC_{D4}$, meaning that counting both Dp and D(6-p) would be redundant.
Central charges of the KK momentum, D0 brane, D2 brane and the fundamental string respectively are then:
\begin{equation}
\Gamma_m\Gamma_7,\hspace{6mm}i\Gamma_7,\hspace{6mm}i\Gamma_{mn},\hspace{6mm}\Gamma_m,
\end{equation}
whose total number adds up to 6+1+15+6=28 as expected.
To connect to the general discussion in section \ref{setup}, we now write the relevant charges used in section \ref{probe}
as $8\times 8$ matrices. For the background D6 brane we get, in a suitable basis:
\begin{equation}
C_A=m_A\Gamma_7
\simeq m_A\begin{psmallmatrix}
&&&& 1 &	&&\\
&&&&& \enskip 1 &&\\
&&&&&	& \enskip 1&\\
&&&&&&& \enskip 1\\
-1&&&&&&&\\
&-1&&&&&&\\
&&-1&&&&	&\\
&&&-1&&&&\\
\end{psmallmatrix}.
\end{equation}
For the D6 probe with fluxes, we get in addition D0, D2, and D4 charges along various planes,
so the central charge is
\begin{equation}
C_B=m_B^{\prime}(\Gamma_7(1+if_1f_2f_3)+\Gamma_{45}(if_1+f_2f_3)+\Gamma_{67}(if_2+f_1f_3)+\Gamma_{89}(if_3+f_1f_2))
\end{equation}
Working in the same basis, this matrix can be brought to the general form given in \eqref{second charge},
with mass $m_B=m_B' \prod_i (1+f_i^2)^{1/2}$ and phases
\begin{align}
e^{2i\phi_1} &= \frac{(1+i f_1)(1-if_2)(1-if_3)}{(1-i f_1)(1+if_2)(1+if_3)}, &\qquad 
e^{2i\phi_3} &= \frac{(1-i f_1)(1-if_2)(1+if_3)}{(1+i f_1)(1+if_2)(1-if_3)}, \\
 e^{2i\phi_2} &= \frac{(1-i f_1)(1+if_2)(1-if_3)}{(1+i f_1)(1-if_2)(1+if_3)},&\qquad
e^{2i\phi_4} &= \frac{(1+i f_1)(1+if_2)(1+if_3)}{(1-i f_1)(1-if_2)(1-if_3)}.
\end{align}
One can see that $\prod_{i=1}^4e^{\phi_i}=1$,
showing how the Pfaffian constraint in eq.~(\ref{C constraints}) is automatically satisfied in this brane setup.

\section{Scalar amplitudes without direct use of superspace} \label{app: amplitudes}

In this appendix, we propose another derivation of the superamplitudes $\mathcal{A}_{\phi\bar{\phi}\eta}$ and $\mathcal{A}_{\phi\bar{\phi}\eta\eta}$ which does not require introducing the massive superspace. We choose the initial scalar states to be identical complex scalars (see the discussion above eq.~\eqref{eq:amp phi}). Now starting from the Ward identities, eqs.~\eqref{ward2} and \eqref{ward1}, we recall:
\begin{equation*}
Q^I=\sum_{i=1}^n q^I_i \qquad {\rm and} \qquad \bar{Q}_I=\sum_{i=1}^n \bar{q}_{I,i} \qquad I={1,2\ldots8}
\end{equation*}
annihilate the superamplitude:
\begin{equation*}
Q^I \mathcal{A}_n=0 \qquad \bar{Q}_I\mathcal{A}_n=0.
\end{equation*}
The goal is to find the linear combinations of supercharges which kill both the scalar and its complex conjugate thus when they act on the amplitude, they only act on the massless particles. In total there are 16 complex supercharges $Q^I_{\alpha}$ and $\bar{Q}_{I\Dot{\alpha}}$, or 32 real supercharges, there are 24 linear combinations of these 32 supercharges which commute and kill the complex scalar state. These linear combinations are different for different scalars in the supermultiplet. We chose our initial scalar state such that it is killed by:
 \begin{equation}
\begin{split}
     q_{\alpha,\phi}^I|\phi\rangle &=0 \quad I=5\ldots8 \\
      \bar{q}_{I\Dot{\alpha},\phi}|\phi\rangle &=0 \quad I=1\ldots4\\
      \left(\bar{q}_{I\Dot{\alpha}}-\frac{p_{\alpha\Dot{\alpha}}}{m}C_{IJ}q^J_{\alpha}\right)|\phi\rangle &=0 \quad I=5\ldots8.
\end{split}
\end{equation}
And for its complex conjugate we have:
\begin{equation}
\begin{split}
   q_{\alpha}^I|\bar{\phi}\rangle &=0 \quad I=1\ldots4 \\
      \bar{q}_{I\Dot{\alpha}}|\bar{\phi}\rangle &=0 \quad I=5\ldots8\\
      \left(\bar{q}_{I\Dot{\alpha}}-\frac{p^{\prime}_{\alpha\Dot{\alpha}}}{m}C_{IJ}q^J_{\alpha}\right)|\bar{\phi}\rangle &=0 \quad I=1\ldots4.
     \end{split}
\end{equation}
It can be easily seen that for a state containing both the scalar and its complex conjugate, there are only 16 independent commuting combinations which kill the product state:
\begin{equation}
\label{supecharges}
    \begin{split}
      \tilde{Q}^I_{\alpha}|\phi\phib\rangle=  \left(\frac{p_{\alpha\Dot{\alpha}}}{m}(\bar{q}_{I\Dot{\alpha},\phi}+\bar{q}_{I\Dot{\alpha},\phib})-\frac{C_{IJ}}{m}(q^J_{\alpha,\phi}+q^J_{\alpha,\phib})\right)|\phi\phib\rangle &=0 \quad I=5\ldots8 \\
       \tilde{Q}^I_{\alpha}|\phi\phib\rangle= \left(\frac{p^{\prime}_{\alpha\Dot{\alpha}}}{m}(\bar{q}_{I\Dot{\alpha},\phi}+\bar{q}_{I\Dot{\alpha},\phib})-\frac{C_{IJ}}{m}(q^J_{\alpha,\phi}+q^J_{\alpha,\phib})\right)|\phi\phib\rangle &=0 \quad I=1\ldots4.
    \end{split}
\end{equation}
Now for any superamplitude containing this product state we have 16 equations which act on the rest of the states in the superamplitude. Schematically:
\begin{equation}
\begin{split}
    \tilde{Q}^I\mathcal{A}_{\eta_1\ldots\eta_n\phi\phib} &=0\\  \left(\tilde{q}^I_{\phi}+\tilde{q}^I_{\phib}+\sum_{i=0}^n \tilde{q}^I_{\eta_i}\right) \mathcal{A}_{\eta_1\ldots\eta_n\phi\phib}&=0 \\
  \sum_{i=0}^n \tilde{q}^I_{\eta_i}\mathcal{A}_{\eta_1\ldots\eta_n\phi\phib}&=0.
    \end{split}
\end{equation}
For the 3-point vertices, we have 8 $\eta_I$s and 16 commuting supercharges acting linearly on the superamplitude, so the problem is overconstrained. 
\begin{equation}
\label{Qs on etas}
\begin{split}
    \tilde{Q^I}\mathcal{A}_{\eta\phi\phib} &=0 \\  \left(\tilde{Q}^I_{\phi}+\tilde{Q}^I_{\bar{\phi}}\right)\mathcal{A}_{\eta\phi\bar{\phi}}+ \tilde{Q}^I_{\eta}\mathcal{A}_{\eta\phi\bar{\phi}}&=0 \\
     \tilde{Q}^I_{\eta}\mathcal{A}_{\eta\phi\phib}&=0.
    \end{split}
\end{equation}
Now using the explicit form of the $\tilde{Q}^I$ given in eq.~\eqref{supecharges} for the left hand side, we have:
\begin{equation}
    \left([\xi|\frac{p_A}{m_A}|3\rangle\eta_I-\frac{C_{IJ}}{m_A}[\xi3]\frac{\partial}{\partial\eta_J}\right)\mathcal{A}^L_{\eta\phi\phib}=0 \qquad I=5\ldots8,
\end{equation}
here $\xi$ is an auxiliary spinor state. Also as expected, this equation is enough to fix the superamplitude. The second equation in \eqref{supecharges} can be used for consistency checks. Solving the first order Grassmannian differential equation we find out:
\begin{equation}
  \mathcal{A}^L_{\phi\phib\eta}=A^L_{\phi\phi\eta}e^{\sum_{I=1}^4\frac{\langle 3|p_A|\xi]C_A^{IJ}}{[3\xi]m_A^2}\eta_I\eta_J},
\end{equation}
which is the covariant form of eq.~\eqref{lhs}. Now let us use eq.~\eqref{Qs on etas} to get the $\mathcal{A}_{\phi\bar{\phi}\eta\eta}$. Here we have 16 superspace parameters, so we need all the 16 supercharges to fix the amplitude.
  \begin{equation}
  \begin{split}
      &\left(\langle 3|\frac{p_A}{m_A}|4]\eta_{3,I}+\langle 4|\frac{p_A}{m_A}|4]\eta_{4,I}-\frac{C_{IJ}}{m_A}[34]\frac{\partial}{\partial\eta_{3,J}}
      \right) \mathcal{A}^L_{\phi\phib\eta \eta}=0 \qquad I=5\ldots8\\
        &\left(\langle 4|\frac{p_A}{m_A}|3]\eta_{4,I}+\langle 3|\frac{p_A}{m_A}|3]\eta_{3,I}-\frac{C_{IJ}}{m_A}[43]\frac{\partial}{\partial\eta_{4,J}}\right) \mathcal{A}^L_{\phi\phib\eta \eta}=0 \qquad I=5\ldots8.
      \end{split}
  \end{equation}
So we get $\mathcal{A}_{\phi\phib \eta \eta}$ to be:
\begin{equation}
\begin{split}
\label{covariant A 2 etas}
 \mathcal{A}^L_{\phi\phib\eta \eta} =& A^L_{\phi\phib\eta \eta} \exp\Bigg(\sum_{I=1}^4\frac{C_A^{IJ}}{[34]m_A}\Big(\langle 3
 |\frac{p_A}{m_A}|4]\eta_{3,I}\eta_{3,J}+\langle 4 |\frac{p_A}{m_A}|4]\eta_{4,I}\eta_{3,J}\\&-\langle 4 |\frac{p_A}{m_A}|3]\eta_{4,I}\eta_{4,J}-\langle 3 |\frac{p_A}{m_A}|3]\eta_{3,I}\eta_{4,J}\Big)\Bigg).
 \end{split}
 \end{equation}
This is the covariant form of eq.~\eqref{AL phi phi eta eta}.

\section{Comments on subtleties with monopole scattering}
\label{app:monopole}

As mentioned in section \ref{amplitude}, there are some subtleties in interpreting the
tree and one-loop scattering amplitudes in eqs.~(\ref{tree pole}) and (\ref{cutintegral}) in the case that  particles have magnetic charges.
These subtleties prevent us from being confident about our extraction of the one-loop
long-distance force between such objects (although the probe limit calculation in section
\ref{probe} does show that there is no precession in this case). 

The difficulty arises when one tries to be fully precise about the meaning of the $t=0$ pole
computed for the tree amplitude in eq.~(\ref{tree pole}).
Imposing $t=0$ and the on-shell conditions for the external particles generally yields \emph{two} complex solutions:
if the incoming particles are along the $z$ axis, these satisfy respectively $(q_x\pm i q_y)=0$.
This is similar to the familiar fact that there are two types of complex on-shell three-particle vertices for massless particles,
but here we are discussing the exchange of a massless particle between two massive ones.
On each of these solutions, the ratio $x_A/x_B$ can be computed to be $e^{\pm \eta_{AB}}$, respectively.
What eq.~(\ref{tree pole}) predicts is really the residue on each of these two separate solutions, which differ by the sign of $\eta_{AB}$.

Now, it is usually assumed that a scattering amplitude for spinless particles is a function only of Mandelstam invariants
$s$ and $t$.
Any such a function will give the same coefficient of $1/t$ for each of the $t=0$ poles, because $s$ depends only on $\cosh(\eta_{AB})$.
However, looking at the prediction (\ref{tree pole}), one finds that the two residues are \emph{not} equal:
\be\begin{aligned}
\mathcal{A}^{\rm tree}_{\phi\phib\phi\phib}\Big|_{\rm pole\ 1}
-\mathcal{A}^{\rm tree}_{\phi\phib\phi\phib}\Big|_{\rm pole\ 2}
&= 8\pi G_N \frac{m_A^2m_B^2}{-t}
\left[\prod_{i=1}^4 2\sinh\left(\frac{\eta-i\phi_I}{2}\right)
-\prod_{i=1}^4 2\sinh\left(\frac{-\eta-i\phi_I}{2}\right) \right]
\\ &= -16\pi i N_{AB} \frac{m_Am_B\sinh\eta_{AB}}{-t} \ .
\end{aligned} \label{pole differences}
\ee
Here we have used that $\prod_{i=1}^4 e^{i\phi_I}=1$ to simplify, and used trigonometric identities to express the discrepancy
in terms of the Dirac-Zwanziger bracket $N_{AB}$ defined in eq.~(\ref{NABdef}).
This is an integer which measures the magnetic flux that effectively couples the two particles.
The conclusion is that when $N_{AB}$ is nonzero, the scattering amplitude cannot be a function of only Mandelstam invariants $s,t$.

This is a well-known result, with a simple semi-classical explanation \cite{Goldhaber:1965cxe,Balian:2005joa}:
the integrated Lorentz force
gives a momentum transfer $\delta \vec{p} \sim N_{AB} \frac{\vec{v}\times \vec{b}}{\vec{b}^2}$,
writing $\vec{b}$ semi-classically as the momentum gradient of a scattering phase requires a strong dependence
on the azimuthal angle, $\mathcal{A}\propto e^{i N_{AB}\varphi}$.
But if the amplitude were to depend only on Mandelstam invariants, it could not possess such a dependence on the azimuthal angle.

A simple resolution is that the $S$-matrix is only expected to be
gauge-invariant under gauge transformations which vanish at infinity.
The vector potential associated with (say) a static monopole with unit charge
depends on the choice of a direction for a ``Dirac string'' $\vec{n}$ which remains visible at arbitrarily large distances.
\begin{equation}
    \vec{A}(r)=\frac{1}{2r}\frac{[\vec{r}\times \vec{n}]}{r-(\vec{r}\cdot\vec{n})}
\end{equation}
Therefore, minimally, the amplitudes must depend in addition on the choice of a Dirac string direction $\vec{n}$.
(The amplitude is really the section of a U(1) bundle and one should avoid the singularity along $\vec{n}$
by gauge-transforming to a different chart.)
For the discussion it is useful to have an explicit form of the gauge transformations which moves the Dirac string.
This transformation must have two singular points on the Riemann sphere,
which is easily constructed by taking the phase of a ratio of spinors (which produce holomorphic functions on the sphere):
\be
 \vec{A}_{n'} = \vec{A}_{n} + \partial_\mu \phi, \qquad \phi = \arg \frac{\langle x n\rangle}{\langle x n'\rangle}.
\ee
Here $\langle x|$ is a spinor in the direction of the coordinate $x$. Since this phase
does not vanish at infinity, it is not surprising that scattering states depend on $\vec{n}$.
Specifically, applying the same gauge transformation to the initial and final states,
the $S$-matrix element will have the following gauge-transformation property:
\be
 \mathcal{A}_{\phi\phib\phi\phib} \to  \mathcal{A}_{\phi\phib\phi\phib}
 \times \left(
\left. \frac{\langle p n\rangle [p'n]}{ [pn]\langle p' n\rangle} \middle/
 \frac{\langle p n'\rangle [p'n']}{ [pn']\langle p' n'\rangle} \right. \right)^{\frac12 N_{AB}}. \label{monopole state transformation}
\ee
So far, our discussion has been nonrelativistic and the spinor $|p\rangle$
is a three-dimensional spinor defined in the rest frame of the system. It admits however a simple relativistically covariant expression.
We simply construct a rank-one $2\times 2$ matrix which has two left-handed chiral indices,
and can therefore be factorized into a product of two spinors:
\be
 \left( \tfrac12[/\!\!\!p_A,/\!\!\!p_B] - \sqrt{(p_A{\cdot}p_B)^2-m_A^2m_B^2}\right)_{\dot\alpha\dot\beta} \equiv |p\rangle_{\dot\alpha} \langle \bar{p}|_{\dot\beta}\, .
\ee
The second spinor, $|\bar{p}\rangle$, points in the direction opposite to $|p\rangle$.
The transformation law then becomes Lorentz-invariant if we label the Dirac string by a null vector $n^\mu$.
The spinor factor in eq.~(\ref{monopole state transformation}) gives the complete dependence of the amplitude on the Dirac string direction;
dividing by a suitable factor we can now remove this dependence.
Insisting to preserve Lorentz and little-group invariance leaves an essentially unique factor:
\be \label{stripped amplitude monopole}
 \mathcal{A}_{\phi\phib\phi\phib} = \left(
 \left.\frac{\langle pn\rangle \langle p' \bar{p}\rangle}{\langle p'n\rangle \langle p\bar{p}\rangle}
\middle/
 \frac{[pn][p'\bar{p}]}{[p'n][p\bar{p}]}\right.
 \right)^{\frac12 N_{AB}} \times \mathcal{A}_{\phi\phib\phi\phib}'(s,t),
\ee
where the reduced amplitude $\mathcal{A}'$ now depends only on Mandelstam invariants.
Pleasingly, in the forward direction, this reproduces the
azimutal angle dependence $(\langle p'\bar{p}\rangle/[p'\bar{p}])^{N/2}\sim e^{i N\phi}$, in agreement with
the semi-classical force calculation. This also shows that the wavefunction is single-valued, provided
$N_{AB}$ is an integer;
from the symmetries we thus expect the nonperturbative amplitude to take the form (\ref{stripped amplitude monopole}).

Although the phase proportional to $N$ looks nonperturbative, we expect physically that the (small)
long-range force between objects can still be computed perturbatively in terms of on-shell exchanges. 
We leave an understanding of cutting rules for monopole objects to the future, with 
the understanding that the expressions in section \ref{oneloopbhscattering} for one-loop cuts are rigorously valid for objects with only electric charges ({\it i.e.}, the two-angle case),
and may or may not be correct for $N_{AB}\neq 0$. 

\section{Quantum mechanical matrix elements}
\label{app:fourier}

Here we compute quantum mechanical matrix elements
of the terms proportional to $\alpha$ in the 0PN and 1PN effective Hamiltonians in eqs.~(\ref{H0PN}) and (\ref{H1PN}):
\be
 H^{(1)} = -\frac{\alpha}{r} + B\frac{\alpha}{\mu^2}\frac{p^2}{r}+ C\frac{\alpha}{\mu^2}\frac{p_r^2}{r}.
\ee
The calculations are standard but it is helpful to have them recorded at one place.
The transform of the first term is familiar,
\be
 \langle p'|\frac{1}{r}|p\rangle = \int d^3r \frac{1}{r} e^{i\vec{r}\cdot(p-p')} = \frac{4\pi}{(p-p')^2}.
\ee
For $p^2/r$, according to the operator ordering stated below eq.~(\ref{H1PN}) we simply get an extra $(p^2+p'^2)/2$,
while for $p_r^2/r$ we use the identity in eq.~(\ref{pr identity}) to reduce it to the computation of:
\be
 \langle p'|r|p\rangle = -\frac{8\pi}{(p-p')^4}.
\ee
In this way we find
\be
 \langle p' | H^{(1)}| p\rangle = -\frac{4\pi \alpha}{(p'-p)^2}
 \left( 1-(B+C) \frac{p'^2+p^2}{2\mu^2}\right)
 +\frac{2\pi \alpha C \hbar^2}{\mu^2} \left(1- \frac{(p'^2-p^2)^2}{(p'-p)^4}\right)\,. \label{H1 matrix element}
\ee
Restricting to the case where $p'^2=p^2$ by energy conservation gives the result in eq.~(\ref{first born}).

In the text we also need the second Born iteration of this potential,
that is the loop integral
\begin{eqnarray}
\int \frac{d^3\ell}{(2\pi)^3} \langle p' |H^{(1)}|\ell \rangle\frac{1}{E_{\ell}-E_p-i0}\langle\ell|H^{(1)}|p\rangle
\label{second born}
\end{eqnarray}
where $E_p = \frac{p^2}{2\mu} + A\frac{p^4}{4\mu^3}$, to be velocity-expanded to 1PN accuracy.
Using that to this order the propagator is
\be
 \frac{1}{\frac{1}{2\mu}(\ell^2-p^2)+\frac{A}{4\mu^3}(\ell^4-p^4)}
 \simeq\frac{2\mu}{\ell^2-p^2}\times \left(1-A\frac{\ell^2+p^2}{2\mu^2}\right),
\ee
the integral becomes
\be\begin{split}
\label{EFTintegrand}
& 2\mu (4\pi \alpha)^2
\left(I_{0}\left(1-(A+2B+2C)\frac{p^2}{\mu^2}\right) - \frac{A+2B+2C}{2\mu^2}I_1 + \frac{C}{\mu^2} (I_3-I_2)\right).
\end{split}\ee
Here $I_{0,1}$ and $I_C$ denote the integrals (using dimensional regularization with
$\bar\mu=\mu(4\pi e^{\gamma_{\rm E}})^{-\eps}$ the $\overline{\rm MS}$ scale):
\be\begin{aligned}
\label{integrals}
I_{0} &\equiv \int \frac{\mu^{2\eps}d^d\ell}{(2\pi)^d} \frac{1}{\big[\ell^2-p^2-i0\big](\ell-p)^2(\ell-p')^2} =
\frac{1}{8\pi} \frac{1}{q^2\sqrt{-p^2-i0}}\left[
\log \frac{q^2}{\bar\mu^2} - \frac{1}{\eps}\right],
\\
I_{1} &\equiv \int \frac{\mu^{2\eps}d^d\ell}{(2\pi)^d} \frac{1}{(\ell-p)^2(\ell-p')^2} = \frac{1}{8q},
\\
I_2 &\equiv \int \frac{\mu^{2\eps}d^d\ell}{(2\pi)^d}\frac{1}{\big[\ell^2-p^2-i0\big](\ell-p)^2}
=-\frac{1}{16\pi \sqrt{-p^2-i0}} \left[ \log \frac{-p^2-i0}{\bar\mu^2}-\frac{1}{\eps}\right],
\\
I_3&\equiv \int \frac{\mu^{2\eps}d^d\ell}{(2\pi)^d}\frac{\ell^2-p^2}{(\ell-p)^4(\ell-p')^2} = \frac12 I_1.
\end{aligned}\ee
Together with the Fourier transform of the second-order potential,
$\langle p'| \frac{1}{r^2}|p\rangle=\frac{2\pi^2}{|p'-p|}$, this gives the result in eq.~(\ref{EFT A1}).

\bibliographystyle{JHEP}
\bibliography{ref}

\providecommand{\href}[2]{#2}\begingroup\raggedright\begin{thebibliography}{10}

\bibitem{Donoghue:1994dn}
J.~F. Donoghue, \emph{{General relativity as an effective field theory: The
  leading quantum corrections}},
  \href{https://doi.org/10.1103/PhysRevD.50.3874}{\emph{Phys. Rev.} {\bfseries
  D50} (1994) 3874--3888},
  [\href{https://arxiv.org/abs/gr-qc/9405057}{{\ttfamily gr-qc/9405057}}].

\bibitem{Burgess:2003jk}
C.~P. Burgess, \emph{{Quantum gravity in everyday life: General relativity as
  an effective field theory}},
  \href{https://doi.org/10.12942/lrr-2004-5}{\emph{Living Rev. Rel.} {\bfseries
  7} (2004) 5--56}, [\href{https://arxiv.org/abs/gr-qc/0311082}{{\ttfamily
  gr-qc/0311082}}].

\bibitem{Porto:2016pyg}
R.~A. Porto, \emph{{The effective field theorist's approach to gravitational
  dynamics}}, \href{https://doi.org/10.1016/j.physrep.2016.04.003}{\emph{Phys.
  Rept.} {\bfseries 633} (2016) 1--104},
  [\href{https://arxiv.org/abs/1601.04914}{{\ttfamily 1601.04914}}].

\bibitem{Buonanno:1998gg}
A.~Buonanno and T.~Damour, \emph{{Effective one-body approach to general
  relativistic two-body dynamics}},
  \href{https://doi.org/10.1103/PhysRevD.59.084006}{\emph{Phys. Rev.}
  {\bfseries D59} (1999) 084006},
  [\href{https://arxiv.org/abs/gr-qc/9811091}{{\ttfamily gr-qc/9811091}}].

\bibitem{Goldberger:2004jt}
W.~D. Goldberger and I.~Z. Rothstein, \emph{{An Effective field theory of
  gravity for extended objects}},
  \href{https://doi.org/10.1103/PhysRevD.73.104029}{\emph{Phys. Rev.}
  {\bfseries D73} (2006) 104029},
  [\href{https://arxiv.org/abs/hep-th/0409156}{{\ttfamily hep-th/0409156}}].

\bibitem{Foffa:2016rgu}
S.~Foffa, P.~Mastrolia, R.~Sturani and C.~Sturm, \emph{{Effective field theory
  approach to the gravitational two-body dynamics, at fourth post-Newtonian
  order and quintic in the Newton constant}},
  \href{https://doi.org/10.1103/PhysRevD.95.104009}{\emph{Phys. Rev.}
  {\bfseries D95} (2017) 104009},
  [\href{https://arxiv.org/abs/1612.00482}{{\ttfamily 1612.00482}}].

\bibitem{Damour:2017ced}
T.~Damour and P.~Jaranowski, \emph{{Four-loop static contribution to the
  gravitational interaction potential of two point masses}},
  \href{https://doi.org/10.1103/PhysRevD.95.084005}{\emph{Phys. Rev.}
  {\bfseries D95} (2017) 084005},
  [\href{https://arxiv.org/abs/1701.02645}{{\ttfamily 1701.02645}}].

\bibitem{Abbott:2016blz}
{\scshape Virgo, LIGO Scientific} collaboration, B.~P. Abbott et~al.,
  \emph{{Observation of Gravitational Waves from a Binary Black Hole Merger}},
  \href{https://doi.org/10.1103/PhysRevLett.116.061102}{\emph{Phys. Rev. Lett.}
  {\bfseries 116} (2016) 061102},
  [\href{https://arxiv.org/abs/1602.03837}{{\ttfamily 1602.03837}}].

\bibitem{Beisert:2010jr}
N.~Beisert et~al., \emph{{Review of AdS/CFT Integrability: An Overview}},
  \href{https://doi.org/10.1007/s11005-011-0529-2}{\emph{Lett. Math. Phys.}
  {\bfseries 99} (2012) 3--32},
  [\href{https://arxiv.org/abs/1012.3982}{{\ttfamily 1012.3982}}].

\bibitem{Gromov:2014caa}
N.~Gromov, V.~Kazakov, S.~Leurent and D.~Volin, \emph{{Quantum spectral curve
  for arbitrary state/operator in AdS$_{5}$/CFT$_{4}$}},
  \href{https://doi.org/10.1007/JHEP09(2015)187}{\emph{JHEP} {\bfseries 09}
  (2015) 187}, [\href{https://arxiv.org/abs/1405.4857}{{\ttfamily 1405.4857}}].

\bibitem{Gurdogan:2015csr}
Ö.~Gürdo?an and V.~Kazakov, \emph{{New Integrable 4D Quantum Field Theories
  from Strongly Deformed Planar $\mathcal N = $ 4 Supersymmetric Yang-Mills
  Theory}}, \href{https://doi.org/10.1103/PhysRevLett.117.201602,
  10.1103/PhysRevLett.117.259903}{\emph{Phys. Rev. Lett.} {\bfseries 117}
  (2016) 201602}, [\href{https://arxiv.org/abs/1512.06704}{{\ttfamily
  1512.06704}}].

\bibitem{Caron-Huot:2014gia}
S.~Caron-Huot and J.~M. Henn, \emph{{Solvable Relativistic Hydrogenlike System
  in Supersymmetric Yang-Mills Theory}},
  \href{https://doi.org/10.1103/PhysRevLett.113.161601}{\emph{Phys. Rev. Lett.}
  {\bfseries 113} (2014) 161601},
  [\href{https://arxiv.org/abs/1408.0296}{{\ttfamily 1408.0296}}].

\bibitem{Alvarez-Jimenez:2018lff}
J.~Alvarez-Jimenez, I.~Cortese, J.~A. García, D.~Gutiérrez-Ruiz and J.~D.
  Vergara, \emph{{Relativistic Runge-Lenz vector: from ${\cal N}=4$ SYM to
  SO(4) scalar field theory}},
  \href{https://arxiv.org/abs/1805.12165}{{\ttfamily 1805.12165}}.

\bibitem{Cremmer:1978ds}
E.~Cremmer and B.~Julia, \emph{{The N=8 Supergravity Theory. 1. The
  Lagrangian}}, \href{https://doi.org/10.1016/0370-2693(78)90303-9}{\emph{Phys.
  Lett.} {\bfseries B80} (1978) 48}.

\bibitem{Cremmer:1979up}
E.~Cremmer and B.~Julia, \emph{{The SO(8) Supergravity}},
  \href{https://doi.org/10.1016/0550-3213(79)90331-6}{\emph{Nucl. Phys.}
  {\bfseries B159} (1979) 141--212}.

\bibitem{Bern:2010ue}
Z.~Bern, J.~J.~M. Carrasco and H.~Johansson, \emph{{Perturbative Quantum
  Gravity as a Double Copy of Gauge Theory}},
  \href{https://doi.org/10.1103/PhysRevLett.105.061602}{\emph{Phys. Rev. Lett.}
  {\bfseries 105} (2010) 061602},
  [\href{https://arxiv.org/abs/1004.0476}{{\ttfamily 1004.0476}}].

\bibitem{Bern:2008qj}
Z.~Bern, J.~J.~M. Carrasco and H.~Johansson, \emph{{New Relations for
  Gauge-Theory Amplitudes}},
  \href{https://doi.org/10.1103/PhysRevD.78.085011}{\emph{Phys. Rev.}
  {\bfseries D78} (2008) 085011},
  [\href{https://arxiv.org/abs/0805.3993}{{\ttfamily 0805.3993}}].

\bibitem{Bern2002}
Z.~Bern, \emph{Perturbative quantum gravity and its relation to gauge theory},
  \href{https://doi.org/10.12942/lrr-2002-5}{\emph{Living Reviews in
  Relativity} {\bfseries 5} (Jul, 2002) 5}.

\bibitem{ArkaniHamed:2008gz}
N.~Arkani-Hamed, F.~Cachazo and J.~Kaplan, \emph{{What is the Simplest Quantum
  Field Theory?}}, \href{https://doi.org/10.1007/JHEP09(2010)016}{\emph{JHEP}
  {\bfseries 09} (2010) 016},
  [\href{https://arxiv.org/abs/0808.1446}{{\ttfamily 0808.1446}}].

\bibitem{Bern:2013uka}
Z.~Bern, S.~Davies, T.~Dennen, A.~V. Smirnov and V.~A. Smirnov,
  \emph{{Ultraviolet Properties of N=4 Supergravity at Four Loops}},
  \href{https://doi.org/10.1103/PhysRevLett.111.231302}{\emph{Phys. Rev. Lett.}
  {\bfseries 111} (2013) 231302},
  [\href{https://arxiv.org/abs/1309.2498}{{\ttfamily 1309.2498}}].

\bibitem{Bern:2018jmv}
Z.~Bern, J.~J. Carrasco, W.-M. Chen, A.~Edison, H.~Johansson, J.~Parra-Martinez
  et~al., \emph{{Ultraviolet Properties of $\mathcal N = 8$ Supergravity at
  Five Loops}},  \href{https://arxiv.org/abs/1804.09311}{{\ttfamily
  1804.09311}}.

\bibitem{Denef:2002ru}
F.~Denef, \emph{{Quantum quivers and Hall / hole halos}},
  \href{https://doi.org/10.1088/1126-6708/2002/10/023}{\emph{JHEP} {\bfseries
  10} (2002) 023}, [\href{https://arxiv.org/abs/hep-th/0206072}{{\ttfamily
  hep-th/0206072}}].

\bibitem{Bates:2003vx}
B.~Bates and F.~Denef, \emph{{Exact solutions for supersymmetric stationary
  black hole composites}},
  \href{https://doi.org/10.1007/JHEP11(2011)127}{\emph{JHEP} {\bfseries 11}
  (2011) 127}, [\href{https://arxiv.org/abs/hep-th/0304094}{{\ttfamily
  hep-th/0304094}}].

\bibitem{Anninos:2012gk}
D.~Anninos, T.~Anous, F.~Denef, G.~Konstantinidis and E.~Shaghoulian,
  \emph{{Supergoop Dynamics}},
  \href{https://doi.org/10.1007/JHEP03(2013)081}{\emph{JHEP} {\bfseries 03}
  (2013) 081}, [\href{https://arxiv.org/abs/1205.1060}{{\ttfamily 1205.1060}}].

\bibitem{Avery:2007xf}
S.~G. Avery and J.~Michelson, \emph{{Mechanics and Quantum Supermechanics of a
  Monopole Probe Including a Coulomb Potential}},
  \href{https://doi.org/10.1103/PhysRevD.77.085001}{\emph{Phys. Rev.}
  {\bfseries D77} (2008) 085001},
  [\href{https://arxiv.org/abs/0712.0341}{{\ttfamily 0712.0341}}].

\bibitem{Elvang:2013cua}
H.~Elvang and Y.-t. Huang, \emph{{Scattering Amplitudes}},
  \href{https://arxiv.org/abs/1308.1697}{{\ttfamily 1308.1697}}.

\bibitem{Hull:1994ys}
C.~M. Hull and P.~K. Townsend, \emph{{Unity of superstring dualities}},
  \href{https://doi.org/10.1016/0550-3213(94)00559-W}{\emph{Nucl. Phys.}
  {\bfseries B438} (1995) 109--137},
  [\href{https://arxiv.org/abs/hep-th/9410167}{{\ttfamily hep-th/9410167}}].

\bibitem{Green:1982sw}
M.~B. Green, J.~H. Schwarz and L.~Brink, \emph{{N=4 Yang-Mills and N=8
  Supergravity as Limits of String Theories}},
  \href{https://doi.org/10.1016/0550-3213(82)90336-4}{\emph{Nucl. Phys.}
  {\bfseries B198} (1982) 474--492}.

\bibitem{Bern:1998ug}
Z.~Bern, L.~J. Dixon, D.~C. Dunbar, M.~Perelstein and J.~S. Rozowsky, \emph{{On
  the relationship between Yang-Mills theory and gravity and its implication
  for ultraviolet divergences}},
  \href{https://doi.org/10.1016/S0550-3213(98)00420-9}{\emph{Nucl. Phys.}
  {\bfseries B530} (1998) 401--456},
  [\href{https://arxiv.org/abs/hep-th/9802162}{{\ttfamily hep-th/9802162}}].

\bibitem{Bern:2008pv}
Z.~Bern, J.~J.~M. Carrasco, L.~J. Dixon, H.~Johansson and R.~Roiban,
  \emph{{Manifest Ultraviolet Behavior for the Three-Loop Four-Point Amplitude
  of N=8 Supergravity}},
  \href{https://doi.org/10.1103/PhysRevD.78.105019}{\emph{Phys. Rev.}
  {\bfseries D78} (2008) 105019},
  [\href{https://arxiv.org/abs/0808.4112}{{\ttfamily 0808.4112}}].

\bibitem{Arkani-Hamed:2017jhn}
N.~Arkani-Hamed, T.-C. Huang and Y.-t. Huang, \emph{{Scattering Amplitudes For
  All Masses and Spins}},  \href{https://arxiv.org/abs/1709.04891}{{\ttfamily
  1709.04891}}.

\bibitem{Guevara:2017csg}
A.~Guevara, \emph{{Holomorphic Classical Limit for Spin Effects in
  Gravitational and Electromagnetic Scattering}},
  \href{https://arxiv.org/abs/1706.02314}{{\ttfamily 1706.02314}}.

\bibitem{Goldhaber:1965cxe}
A.~S. Goldhaber, \emph{{Role of Spin in the Monopole Problem}},
  \href{https://doi.org/10.1103/PhysRev.140.B1407}{\emph{Phys. Rev.} {\bfseries
  140} (1965) B1407--B1414}.

\bibitem{Balian:2005joa}
Y.~M. Shnir, \emph{{Magnetic Monopoles}}.
\newblock Text and Monographs in Physics. Springer, Berlin/Heidelberg, 2005,
  \href{https://doi.org/10.1007/3-540-29082-6}{10.1007/3-540-29082-6}.

\bibitem{Holstein:2016fxh}
B.~R. Holstein, \emph{{Analytical On-shell Calculation of Higher Order
  Scattering: Massive Particles}},
  \href{https://arxiv.org/abs/1610.07957}{{\ttfamily 1610.07957}}.

\bibitem{BjerrumBohr:2006yw}
N.~E.~J. Bjerrum-Bohr, D.~C. Dunbar, H.~Ita, W.~B. Perkins and K.~Risager,
  \emph{{The No-Triangle Hypothesis for N=8 Supergravity}},
  \href{https://doi.org/10.1088/1126-6708/2006/12/072}{\emph{JHEP} {\bfseries
  12} (2006) 072}, [\href{https://arxiv.org/abs/hep-th/0610043}{{\ttfamily
  hep-th/0610043}}].

\bibitem{Denner:1991qq}
A.~Denner, U.~Nierste and R.~Scharf, \emph{{A Compact expression for the scalar
  one loop four point function}},
  \href{https://doi.org/10.1016/0550-3213(91)90011-L}{\emph{Nucl. Phys.}
  {\bfseries B367} (1991) 637--656}.

\bibitem{Feinberg:1988yw}
G.~Feinberg and J.~Sucher, \emph{{The Two Photon Exchange Force Between Charged
  Systems. 1. Spinless Particles}},
  \href{https://doi.org/10.1103/PhysRevD.38.3763,
  10.1103/PhysRevD.44.3997}{\emph{Phys. Rev.} {\bfseries D38} (1988) 3763}.

\bibitem{BjerrumBohr:2002kt}
N.~E.~J. Bjerrum-Bohr, J.~F. Donoghue and B.~R. Holstein, \emph{{Quantum
  gravitational corrections to the nonrelativistic scattering potential of two
  masses}}, \href{https://doi.org/10.1103/PhysRevD.71.069903,
  10.1103/PhysRevD.67.084033}{\emph{Phys. Rev.} {\bfseries D67} (2003) 084033},
  [\href{https://arxiv.org/abs/hep-th/0211072}{{\ttfamily hep-th/0211072}}].

\bibitem{Brambilla:2004jw}
N.~Brambilla, A.~Pineda, J.~Soto and A.~Vairo, \emph{{Effective field theories
  for heavy quarkonium}},
  \href{https://doi.org/10.1103/RevModPhys.77.1423}{\emph{Rev. Mod. Phys.}
  {\bfseries 77} (2005) 1423},
  [\href{https://arxiv.org/abs/hep-ph/0410047}{{\ttfamily hep-ph/0410047}}].

\bibitem{Donoghue:1995cz}
J.~F. Donoghue, \emph{{Introduction to the effective field theory description
  of gravity}},  in \emph{{Advanced School on Effective Theories Almunecar,
  Spain, June 25-July 1, 1995}}, 1995,
  \href{https://arxiv.org/abs/gr-qc/9512024}{{\ttfamily gr-qc/9512024}}.

\bibitem{Iwasaki:1971vb}
Y.~Iwasaki, \emph{{Quantum theory of gravitation vs. classical theory. -
  fourth-order potential}},
  \href{https://doi.org/10.1143/PTP.46.1587}{\emph{Prog. Theor. Phys.}
  {\bfseries 46} (1971) 1587--1609}.

\bibitem{Hiida:1972xs}
K.~Hiida and H.~Okamura, \emph{{Gauge transformation and gravitational
  potentials}}, \href{https://doi.org/10.1143/PTP.47.1743}{\emph{Prog. Theor.
  Phys.} {\bfseries 47} (1972) 1743--1757}.

\bibitem{Bernard:2016wrg}
L.~Bernard, L.~Blanchet, A.~Bohé, G.~Faye and S.~Marsat, \emph{{Energy and
  periastron advance of compact binaries on circular orbits at the fourth
  post-Newtonian order}},
  \href{https://doi.org/10.1103/PhysRevD.95.044026}{\emph{Phys. Rev.}
  {\bfseries D95} (2017) 044026},
  [\href{https://arxiv.org/abs/1610.07934}{{\ttfamily 1610.07934}}].

\bibitem{Herzog:2009fw}
C.~P. Herzog and T.~Klose, \emph{{The Perfect Atom: Bound States of
  Supersymmetric Quantum Electrodynamics}},
  \href{https://doi.org/10.1016/j.nuclphysb.2010.06.004}{\emph{Nucl. Phys.}
  {\bfseries B839} (2010) 129--156},
  [\href{https://arxiv.org/abs/0912.0733}{{\ttfamily 0912.0733}}].

\bibitem{Cheung:2018wkq}
C.~Cheung, I.~Z. Rothstein and M.~P. Solon, \emph{{From Scattering Amplitudes
  to Classical Potentials in the Post-Minkowskian Expansion}},
  \href{https://arxiv.org/abs/1808.02489}{{\ttfamily 1808.02489}}.

\bibitem{Bjerrum-Bohr:2018xdl}
N.~E.~J. Bjerrum-Bohr, P.~H. Damgaard, G.~Festuccia, L.~Planté and P.~Vanhove,
  \emph{{General Relativity from Scattering Amplitudes}},
  \href{https://arxiv.org/abs/1806.04920}{{\ttfamily 1806.04920}}.

\bibitem{Damour:2016gwp}
T.~Damour, \emph{{Gravitational scattering, post-Minkowskian approximation and
  Effective One-Body theory}},
  \href{https://doi.org/10.1103/PhysRevD.94.104015}{\emph{Phys. Rev.}
  {\bfseries D94} (2016) 104015},
  [\href{https://arxiv.org/abs/1609.00354}{{\ttfamily 1609.00354}}].

\bibitem{Nabet:2014kva}
B.~M. Nabet and B.~Kol, \emph{{Leading anomalies, the drift Hamiltonian and the
  relativistic two-body system}},
  \href{https://arxiv.org/abs/1408.2628}{{\ttfamily 1408.2628}}.

\bibitem{Damour:2008qf}
T.~Damour, P.~Jaranowski and G.~Schaefer, \emph{{Effective one body approach to
  the dynamics of two spinning black holes with next-to-leading order
  spin-orbit coupling}},
  \href{https://doi.org/10.1103/PhysRevD.78.024009}{\emph{Phys. Rev.}
  {\bfseries D78} (2008) 024009},
  [\href{https://arxiv.org/abs/0803.0915}{{\ttfamily 0803.0915}}].

\bibitem{Aharony:1999ti}
O.~Aharony, S.~S. Gubser, J.~M. Maldacena, H.~Ooguri and Y.~Oz, \emph{{Large N
  field theories, string theory and gravity}},
  \href{https://doi.org/10.1016/S0370-1573(99)00083-6}{\emph{Phys. Rept.}
  {\bfseries 323} (2000) 183--386},
  [\href{https://arxiv.org/abs/hep-th/9905111}{{\ttfamily hep-th/9905111}}].

\bibitem{Bloch:1937pw}
F.~Bloch and A.~Nordsieck, \emph{{Note on the Radiation Field of the
  electron}}, \href{https://doi.org/10.1103/PhysRev.52.54}{\emph{Phys. Rev.}
  {\bfseries 52} (1937) 54--59}.

\bibitem{Strominger:2017zoo}
A.~Strominger, \emph{{Lectures on the Infrared Structure of Gravity and Gauge
  Theory}},  \href{https://arxiv.org/abs/1703.05448}{{\ttfamily 1703.05448}}.

\bibitem{Drummond:2008vq}
J.~M. Drummond, J.~Henn, G.~P. Korchemsky and E.~Sokatchev, \emph{{Dual
  superconformal symmetry of scattering amplitudes in N=4 super-Yang-Mills
  theory}}, \href{https://doi.org/10.1016/j.nuclphysb.2009.11.022}{\emph{Nucl.
  Phys.} {\bfseries B828} (2010) 317--374},
  [\href{https://arxiv.org/abs/0807.1095}{{\ttfamily 0807.1095}}].

\bibitem{deAzcarraga:1989mza}
J.~A. de~Azcarraga, J.~P. Gauntlett, J.~M. Izquierdo and P.~K. Townsend,
  \emph{{Topological Extensions of the Supersymmetry Algebra for Extended
  Objects}}, \href{https://doi.org/10.1103/PhysRevLett.63.2443}{\emph{Phys.
  Rev. Lett.} {\bfseries 63} (1989) 2443}.

\bibitem{Polchinski:1998rr}
J.~Polchinski, \emph{{String theory. Vol. 2: Superstring theory and beyond}}.
\newblock Cambridge Monographs on Mathematical Physics. Cambridge University
  Press, 2007,
  \href{https://doi.org/10.1017/CBO9780511618123}{10.1017/CBO9780511618123}.

\end{thebibliography}\endgroup
\end{document}